\documentclass[extra, mreferee]{gji}
\usepackage{epsfig, color}
\usepackage{amssymb, amsmath}
\usepackage{setspace}

\DeclareMathOperator{\atantwo}{atan2}

\title[Green's functions in 2D time-dependent materials]
 {Green's functions, propagation invariants, reciprocity theorems, wave-field representations and propagator matrices in 2D time-dependent materials}

\author[Wapenaar]
{\small Kees Wapenaar\\
Department of Geoscience and Engineering, Delft University of Technology, The Netherlands}
\pagerange{1--99}
\volume{999}
\pubyear{2024}

\begin{document}
\begin{spacing}{2.0}
\label{firstpage}

\maketitle

\begin{summary}

{
The study of wave propagation and scattering in time-dependent materials is a rapidly growing field of research. Whereas for 1D applications there is a simple relation between
the wave equations for space-dependent and time-dependent materials, this relation is less straightforward for multi-dimensional materials.

This paper discusses fundamental aspects of 2D electromagnetic and acoustic wave propagation and scattering in homogeneous, time-dependent materials.
This encompasses a review of transmission and reflection at a single time boundary, a discussion of the Green's function and its symmetry properties in a piecewise continuous time-dependent material,
a discussion of propagation invariants (including the net field-momentum density), general reciprocity theorems, and wave field representations. 

Analogous to the well-known expression for Green's function retrieval by time-correlation of passive measurements in a space-dependent material, 
an expression is derived for Green's function retrieval by space-correlation of passive measurements in a time-dependent material.

The paper concludes with the discussion of the propagator matrix for a piecewise continuous time-dependent material, its symmetry properties and its relation with the Green's function.
}
 \end{summary}


\section{Introduction}

Temporal changes of material parameters have an effect on wave propagation and scattering, comparable with, but not identical to, the effect of spatial changes.
Since the initial work on wave propagation and scattering in time-dependent materials  \citep{Morgenthaler58IRE, Jiang75IEEE}, 
the research in this field has recently got significant momentum as a result of advances in the engineering of dynamic metamaterials \citep{Caloz2020IEEE1}.
Most applications deal with electromagnetic waves \citep{Koutserimpas2018IEEE, Mounaix2020NC, Ramaccia2020OL, Garnier2021SIAM, Apffel2022PRL, Moussa2023NP, Ptitcyn2023IEEE}, 
but mechanical wave propagation 
and  scattering in time-dependent materials is also a rapidly emerging research area. It is proposed as an alternative way of doing ultrasonic time-reversal experiments
\citep{Bacot2016NP, Fink2017EPJ,  Bal2019SIAM, Peng2020JPC, Hidalgo2023PRL} 
and its potential use in seismic imaging and monitoring is being investigated \citep{Innanen2018SEG, Huang2022IMAGE}.

Various authors discuss the analogy between the wave equations for time-dependent and space-dependent materials 
\citep{Mendonca2002PS, Xiao2014OL, Hoop2014WM, Salem2015arXiv, Torrent2018PRB, Caloz2020IEEE2, Manen2024arXiv}. 
For 1D applications, the roles of the time- and space-coordinates are interchanged between the wave equations for both types of material
\citep{Salem2015arXiv, Torrent2018PRB, Caloz2020IEEE2, Manen2024arXiv}. 
However, causality conditions apply to the time coordinate in both types of material and hence they are not interchanged. 
As a consequence, solutions of the wave equation in a time-dependent material are, in general, not one-to-one exchangeable with those in a space-dependent material.
In a recent paper \citep{Wapenaar2024WM} we systematically analyze the similarities and differences of waves in 1D space-dependent and time-dependent materials.
In multi-dimensional materials, the number of space dimensions (two or three) is different from the number of time dimensions (one) and hence the
analogy between the wave equations for both types of material is less straightforward. 

In the current paper we discuss fundamental aspects of wave propagation and scattering in 2D time-dependent materials. 
The discussion partly reviews work discussed in the references mentioned above but also contains new results. Like in our paper on 1D materials  \citep{Wapenaar2024WM},
we use a unified notation which simultaneously captures electromagnetic, acoustic and elastodynamic shear waves.
We discuss transmission and reflection coefficients, Green's functions, propagation invariants, reciprocity theorems, 
wave field representations, Green's function retrieval, propagator matrices and the relation between the Green's function
and the propagator matrix of a time-dependent material.

\section{Unified wave equation}\label{sec2}

We consider 2D wave propagation in the $x,z$-plane, assuming the material parameters, the sources and the wave fields are independent of the $y$-coordinate. 
We denote position in this plane by the Cartesian coordinate vector ${\bf x}=(x,z)$ and time by $t$.
We capture electromagnetic waves (transverse electric and transverse magnetic), acoustic waves and horizontally polarized shear waves by the following unified equations
\citep{Carcione95WM, Hoop95Book, Wapenaar2001RS, Carcione2002SGG, Burns2020NJP}
\begin{eqnarray}
\partial_t U + {\bf \nabla}\cdot {\bf Q} &=& a,\label{eq1}\\
\partial_t {\bf V} + {\bf \nabla} P &=& {\bf b},\label{eq2}
\end{eqnarray}
where $\partial_t$ stands for the partial differential operator $\frac{\partial}{\partial t}$ and ${\bf \nabla}$ is the 2D nabla operator, defined as ${\bf \nabla}=(\partial_x, \partial_z)$.
Furthermore, $U({\bf x},t)$, ${\bf V}({\bf x},t)$, $P({\bf x},t)$ and ${\bf Q}({\bf x},t)$ are space- and time-dependent wave-field quantities 
and $a({\bf x},t)$ and ${\bf b}({\bf x},t)$ are space- and time-dependent source quantities. The boldface quantities ${\bf V}$, ${\bf Q}$ and ${\bf b}$ 
denote vectors with two components, hence ${\bf V}=(V_x,V_z)$, etc. 
All quantities are further specified in Table \ref{table1} for the different wave phenomena considered in this paper.
For electromagnetic waves, equations (\ref{eq1}) and (\ref{eq2}) are Maxwell's equations in the $x,z$-plane  \citep{Landau60Book, Hoop95Book}.
For example, for  transverse electric waves (row 1 in Table \ref{table1}) we have ${\bf Q}=(H_z,-H_x)$, hence, 
${\bf \nabla}\cdot {\bf Q}=\partial_xH_z -\partial_zH_x=-({\bf \nabla}\times{\bf H})_y$, 
which is minus the $y$-component of the curl of the magnetic field vector ${\bf H}({\bf x},t)$.
Hence, equation (\ref{eq1}) reads for this situation $\partial_tD_y-({\bf \nabla}\times{\bf H})_y=-J_y^{\rm e}$. 
With some reordering, equation (\ref{eq2}) can be shown to be the 2D version of $\partial_t{\bf B}+{\bf \nabla}\times{\bf E}=-{\bf J}^m$.
In a similar way it can be shown that for transverse magnetic waves (row 2 in Table \ref{table1}), equations (\ref{eq1}) and (\ref{eq2}) are again Maxwell's equations in the $x,z$-plane, but in reversed order.
For acoustic waves (row 3 in Table \ref{table1}) we have ${\bf \nabla}\cdot {\bf Q}=\partial_xv_x+ \partial_zv_z$, which is the divergence of the 
particle velocity vector ${\bf v}({\bf x},t)$. Hence, for this situation equation (\ref{eq1}) stands for $-\partial_t\Theta + {\bf \nabla}\cdot {\bf v}=q$, which is the acoustic deformation equation,
and equation (\ref{eq2}) becomes $\partial_t{\bf m}+{\bf \nabla}p={\bf f}$, which quantifies equilibrium of momentum  \citep{Torrent2018PRB,Hoop95Book}. 
Similarly, for horizontally polarized shear waves (row 4 in Table \ref{table1}), equation (\ref{eq1}) stands for equilibrium of momentum and equation (\ref{eq2}) is the 2D elastic deformation equation.

\begin{table}
\caption{
Specification of the quantities in equations (\ref{eq1}) -- (\ref{eq4}). For TE (transverse electric) and TM (transverse magnetic) waves, the
quantities are electric and magnetic flux densities ${D}$ and ${B}$,
electric and magnetic field strengths ${E}$ and ${H}$, permittivity $\varepsilon$, permeability $\mu$,
and external electric and magnetic current densities ${J}^{\rm e}$ and ${J}^{\rm m}$. 
For acoustic waves they are  dilatation $\Theta$, mechanical momentum density ${m}$, acoustic pressure $p$, particle velocity ${v}$,  
compressibility $\kappa$, mass density $\rho$,  volume-injection rate density $q$ and external force density ${f}$.
For SH (horizontally polarised shear) waves, $m$, $v$, $\rho$ and $f$ are defined the same as for acoustic waves, and 
the additional quantities are strain $e$,  stress $\tau$, shear modulus $\mu$ and external  deformation rate density $h$.}\label{table1}
\begin{tabular}
{lccccccccccc}
\hline
& $U$ & $V_x$ &$V_z$ &$P$ & $Q_x$ &$Q_z$ &$\alpha$ &$\beta$  & $a$ & $b_x$ &$b_z$  \\
\hline
1. TE   & $D_y$ & $B_z$ &$-B_x$& $E_y$ & $H_z$  & $-H_x$ &$\varepsilon$ &$\mu$  &  $-J_y^{\rm e}$ & $-J_z^{\rm m}$ &$J_x^{\rm m}$  \\
2. TM   & $B_y$ & $-D_z$ &$D_x$& $H_y$ & $-E_z$ &$E_x$ &$\mu$ &$\varepsilon$  & $-J_y^{\rm m}$ & $J_z^{\rm e}$&$-J_x^{\rm e}$  \\
3. Acoustic  & $-\Theta$ & $m_x$ &$m_z$& $p$ & $v_x$ & $v_z$ &$\kappa$ &$\rho$  & $q$ & $f_x$ &$f_z$ \\
4. SH  & $m_y$ & $-2e_{yx}$ & $-2e_{yz}$& $v_y$ & $-\tau_{yx}$ & $-\tau_{yz}$ &$\rho$ &$\frac{1}{\mu}$ & $f_y$ & $2h_{yx}$ &$2h_{yz}$  \\
\hline
\end{tabular}
\end{table}

The wave-field quantities are mutually related via the  constitutive equations, as follows 
\begin{eqnarray}
U &=& \alpha P,\label{eq3}\\
{\bf V} &=& \beta {\bf Q},\label{eq4}
\end{eqnarray}
where $\alpha$ and $\beta$ are material parameters. 
They are also specified in Table \ref{table1} for the different wave phenomena. 
We can use these  equations  to eliminate two of the four wave field quantities from equations (\ref{eq1}) and (\ref{eq2}). We consider three cases.

\begin{enumerate}
\item {\bf Inhomogeneous, time-dependent material, with parameters $\alpha({\bf x},t)$ and $\beta({\bf x},t)$.}\\
Substituting equations (\ref{eq3}) and (\ref{eq4}) straightforwardly into equations (\ref{eq1}) and (\ref{eq2}) yields
\begin{eqnarray}
\partial_t(\alpha P) + {\bf \nabla}\cdot {\bf Q} &=& a,\label{eq1ag}\\
\partial_t(\beta {\bf Q}) + {\bf \nabla} P &=& {\bf b}.\label{eq2ag}
\end{eqnarray}
For example, for the special situation of acoustic waves (row 3 in Table \ref{table1}) this gives  $\partial_t(\kappa p) + {\bf \nabla}\cdot {\bf v}=q$
and $\partial_t(\rho{\bf v})+{\bf \nabla}p={\bf f}$ \citep{Nassar2017JMPS}.\\

\item {\bf Inhomogeneous, time-independent material,  with parameters $\alpha({\bf x})$ and $\beta({\bf x})$.}\\
For this situation we have
$\partial_t U=\partial_t(\alpha P)=\alpha\partial_tP$ and 
$\partial_t{\bf V}=\partial_t(\beta {\bf Q})=\beta\partial_t{\bf Q}$. 
Hence, we obtain from equations (\ref{eq1ag}) and (\ref{eq2ag})
\begin{eqnarray}
\alpha\partial_t P +{\bf \nabla}\cdot{\bf Q} &=& a,\label{eq6}\\
\beta\partial_t {\bf Q} + {\bf \nabla} P &=& {\bf b}.\label{eq7}
\end{eqnarray}
As an example, for acoustic waves these expressions become $\kappa\partial_tp + {\bf \nabla}\cdot {\bf v}=q$ and $\rho\partial_t{\bf v}+{\bf \nabla}p={\bf f}$.

The well-known system of equations (\ref{eq6}) and (\ref{eq7}) for the wave field quantities $P$ and ${\bf Q}$ underlies wave propagation in inhomogeneous, time-independent materials
with parameters $\alpha({\bf x})$ and $\beta({\bf x})$, which may vary continuously with space.
When the material contains space boundaries with normal ${\bf n}({\bf x})$, equations (\ref{eq6}) and (\ref{eq7}) are supplemented with boundary conditions.
The boundary conditions state that $P$ and ${\bf Q}\cdot{\bf n}$
 are continuous over those space boundaries.\\

\item {\bf Homogeneous, time-dependent material, with parameters  $\alpha(t)$ and $\beta(t)$.}\\
For this situation we cannot use equations (\ref{eq6}) and (\ref{eq7}).
Instead, using equations (\ref{eq3}) and (\ref{eq4}), we now have
${\bf \nabla}P={\bf \nabla}(\frac{1}{\alpha}U)=\frac{1}{\alpha}{\bf \nabla}U$ and ${\bf \nabla}\cdot{\bf Q}={\bf \nabla}\cdot(\frac{1}{\beta}{\bf V})=\frac{1}{\beta}{\bf \nabla}\cdot{\bf V}$.
Hence,  elimination of $P$ and ${\bf Q}$ from equations (\ref{eq1}) and (\ref{eq2}) yields
\begin{eqnarray}
\partial_t U + \frac{1}{\beta}{\bf \nabla}\cdot {\bf V} &=& a,\label{eq49}\\
\partial_t {\bf V} + \frac{1}{\alpha}{\bf \nabla} U &=& {\bf b}.\label{eq50}
\end{eqnarray}
For example, for acoustic waves these expressions read $-\partial_t\Theta + \frac{1}{\rho}{\bf \nabla}\cdot {\bf m}=q$
and $\partial_t{\bf m}-\frac{1}{\kappa}{\bf \nabla}\Theta={\bf f}$.

The system of equations (\ref{eq49}) and (\ref{eq50}) for the wave field quantities $U$ and ${\bf V}$ underlies wave propagation in homogeneous, 
time-dependent materials with parameters $\alpha(t)$ and $\beta(t)$, which may vary continuously with time.
When the material contains time boundaries, equations (\ref{eq49}) and (\ref{eq50}) are supplemented with boundary conditions. 
The boundary conditions state that $U$ and ${\bf V}$
are continuous over those time boundaries
\citep{Xiao2014OL, Morgenthaler58IRE, Mendonca2002PS, Caloz2020IEEE2, Hoop2014WM}.
\end{enumerate}

Although equations (\ref{eq1ag}) and (\ref{eq2ag}) hold for materials that are simultaneously space- and time-dependent, we restrict the discussion in this paper to homogeneous 
 time-dependent materials, governed by equations (\ref{eq49}) and (\ref{eq50}), supplemented with boundary conditions for $U$ and ${\bf V}$. This allows the translation of several fundamental 
 concepts known for waves in inhomogeneous time-independent materials (case ii) to waves in homogeneous time-dependent materials (case iii).
For aspects of wave propagation in materials that are simultaneously space- and time-dependent (case i), see references 
\cite{Caloz2020IEEE2, Apffel2022PRL, Manen2024arXiv, Ammari2024RS}.
For non-reciprocal wave propagation in metamaterials, we refer to 
\cite{Lurie2007Book, Willis2011RS, Willis2012CRM, Trainiti2016NJP, Haberman2016PhysicsToday, Nassar2017JMPS, Nassar2017RS, Goldsberry2019JASA, Sotoodehfar2023OE}.

Next to the first-order equations (\ref{eq49}) and (\ref{eq50}), it
 is also useful to consider a second order wave equation for the scalar field $U({\bf x},t)$. This equation is obtained by eliminating ${\bf V}({\bf x},t)$  from equations (\ref{eq49}) and (\ref{eq50}), 
which yields 
\begin{eqnarray}
\partial_t(\beta\partial_tU) - \beta c^2{\bf \nabla}^2U = s,\label{eq31a}
\end{eqnarray}
with  $c(t)$ being the time-dependent propagation velocity, defined as
\begin{eqnarray}
c=\frac{1}{\sqrt{\alpha\beta}},\label{eq8a}
\end{eqnarray}
$s({\bf x},t)$ being the source function, defined as
\begin{eqnarray}
s=\partial_t(\beta a) - {\bf \nabla}\cdot {\bf b},\label{eqsource}
\end{eqnarray}
and ${\bf \nabla}^2={\bf \nabla}\cdot{\bf \nabla}=\partial_x^2+\partial_z^2$. 
Note that, unlike in the wave equation for an inhomogeneous time-independent material, in which a material parameter appears between the space-derivatives,
in equation (\ref{eq31a}) a material parameter appears between the time-derivatives \citep{Zurita2009PRA, Torrent2018PRB}.
Finally, we define a time-dependent quantity $\eta(t)$ as
\begin{eqnarray}
\eta=\sqrt{\frac{\beta}{\alpha}}=\beta c=\frac{1}{\alpha c}.
\end{eqnarray}
For TE and acoustic waves (rows 1 and 3 in Table \ref{table1}), $\eta$ stands for impedance, whereas it stands for admittance
for TM and SH waves (rows 2 and 4 in Table \ref{table1}).

\section{Transmission and reflection of plane waves at a time boundary}

\subsection{Transmission and reflection coefficients}

We review transmission and reflection coefficients for a monochromatic plane wave that is incident on a time boundary 
\citep{Xiao2014OL, Morgenthaler58IRE, Mendonca2002PS, Caloz2020IEEE2}. 
For convenience we define the time boundary at $t=0$. 
For this analysis, the medium parameters $\alpha(t)$ and $\beta(t)$ are step functions of time, with a finite jump at $t=0$.
For $t<0$ we have time-independent parameters $\alpha_1$ and $\beta_1$. 
Similarly, for $t>0$ we have time-independent parameters $\alpha_2$ and $\beta_2$.
For $t<0$, the incident monochromatic plane wave is described in complex notation by
\begin{eqnarray}
U_I({\bf x},t)&=&\exp\{i({\bf k}_1\cdot{\bf x}-\omega_1t)\},\label{eq3031}
\end{eqnarray}
where subscript $I$ stands for incident, $i$ is the imaginary unit,
$\omega_1$ is the angular frequency of the incident wave and ${\bf k}_1$ is its wave vector, defined as ${\bf k}_1=(k_{x,1},k_{z,1})$. 
For convenience we consider a unit amplitude and zero phase at ${\bf x}={\bf 0}$ and $t=0$ (where ${\bf 0}=(0,0)$). 
The propagation angle (relative to the $z$-axis) is equal to $\atantwo(k_{x,1},k_{z,1})$. 
According to equation (\ref{eq31a}) for the source-free situation (i.e., for $s=0$), 
$U_I({\bf x},t)$ should obey $\partial_t^2U_I=c_1^2{\bf \nabla}^2U_I$, with $c_1=1/\sqrt{\alpha_1\beta_1}$.
This implies $\omega_1^2=c_1^2|{\bf k}_1|^2$ or, choosing positive square-roots on both sides, $\omega_1=c_1|{\bf k}_1|$. 

The boundary conditions state that $U$ and ${\bf V}$ are continuous over the time boundary
\citep{Xiao2014OL, Morgenthaler58IRE, Mendonca2002PS, Caloz2020IEEE2, Hoop2014WM}, hence, we also need an expression for ${\bf V}_I$.
From equations (\ref{eq50}) (with ${\bf b}={\bf 0})$ and (\ref{eq3031}) we obtain
\begin{eqnarray}
{\bf V}_I({\bf x},t)&=&\eta_1\frac{{\bf k}_1}{|{\bf k}_1|}\exp\{i({\bf k}_1\cdot{\bf x}-\omega_1t)\},
\end{eqnarray}
with $\eta_1=1/(\alpha_1c_1)$.
Unlike in the situation of scattering at a space boundary, where the reflected wave propagates in the same half-space as the incident wave, 
in the situation of a time boundary there is no reflected wave before the time boundary,
 since this would violate causality. The reflected and transmitted waves both exist only after the time boundary.
For $t>0$, the transmitted monochromatic plane wave is described by
\begin{eqnarray}
U_T({\bf x},t)&=&T_u\exp\{i({\bf k}_2\cdot{\bf x}-\omega_2t)\},\\
{\bf V}_T({\bf x},t)&=&T_u\eta_2
\frac{{\bf k}_2}{|{\bf k}_2|}\exp\{i({\bf k}_2\cdot{\bf x}-\omega_2t)\},
\end{eqnarray}
with  
$T_u$ denoting the transmission coefficient for the wave-field quantity $U$. Furthermore, 
$\omega_2$ and ${\bf k}_2=(k_{x,2},k_{z,2})$ are the angular frequency and wave vector of the transmitted wave.
They are related via $\omega_2=c_2|{\bf k}_2|$, with $c_2=1/\sqrt{\alpha_2\beta_2}$. Finally, $\eta_2=1/(\alpha_2c_2)$.
The reflected wave for $t>0$ is described by
\begin{eqnarray}
U_R({\bf x},t)&=&R_u\exp\{i({\bf k}_2\cdot{\bf x}+\omega_2t)\},\\
{\bf V}_R({\bf x},t)&=&
-R_u\eta_2\frac{{\bf k}_2}{|{\bf k}_2|}\exp\{i({\bf k}_2\cdot{\bf x}+\omega_2t)\},\label{eq3036}
\end{eqnarray}
with $R_u$ denoting the reflection coefficient for the wave-field quantity $U$. 
Note the opposite signs in front of the terms $\omega_2t$ for the transmitted and reflected waves, which implies these waves propagate in opposite directions.
Since $U({\bf x},t)$ and ${\bf V}({\bf x},t)$ are continuous at $t=0$, we have
\begin{eqnarray}
U_I({\bf x},0)&=&U_T({\bf x},0)+U_R({\bf x},0),\label{eq310}\\
{\bf V}_I({\bf x},0)&=&{\bf V}_T({\bf x},0)+{\bf V}_R({\bf x},0).\label{eq311}
\end{eqnarray}
These equations should hold for all ${\bf x}$, which implies ${\bf k}_1={\bf k}_2$. 
From this it follows that the angular frequency of the transmitted and reflected waves is different from that of the incident wave, according to
$\omega_2=\frac{c_2}{c_1}\omega_1$ \citep{Apffel2022PRL,Mendonca2002PS, Caloz2020IEEE2}. In other words, frequency conversion occurs at a time boundary.
Substituting the appropriate expressions into equations (\ref{eq310}) and (\ref{eq311}) yields
\begin{eqnarray}
1&=&T_u+R_u,\label{eq48a}\\
\eta_1&=&\eta_2(T_u-R_u).\label{eq48b}
\end{eqnarray}
From this we obtain the following expressions for the  transmission and reflection coefficients 
\begin{eqnarray}
T_u&=&\frac{\eta_2+\eta_1}{2\eta_2},\label{eq90312}\\
R_u&=&\frac{\eta_2-\eta_1}{2\eta_2}.\label{eq90311}
\end{eqnarray}
Note that, unlike for the situation of a space boundary, these coefficients are independent of the wave vector and hence of the propagation angle.

By multiplying the left- and right-hand sides of equation (\ref{eq48a}) with those of equation (\ref{eq48b}), we obtain the following relation between the transmission and reflection coefficients
\begin{eqnarray}
\eta_1=\eta_2(T_u^2-R_u^2).\label{eq24u}
\end{eqnarray}
Finally, we define  transmission and reflection coefficients $T_p$ and $R_p$ for the wave-field quantity $P$. Using equation (\ref{eq3}), we obtain  
$T_p=\frac{\alpha_1}{\alpha_2}T_u$ and $R_p=\frac{\alpha_1}{\alpha_2}R_u$, or
\begin{eqnarray}
T_p&=&\frac12\biggl(\frac{\alpha_1}{\alpha_2}+\frac{c_2}{c_1}\biggr),\\
R_p&=&\frac12\biggl(\frac{\alpha_1}{\alpha_2}-\frac{c_2}{c_1}\biggr).
\end{eqnarray}
These expressions are consistent with those given by  \cite{Xiao2014OL}.

\subsection{Conservation of net field-momentum density}\label{sec3b}

For electromagnetic waves, we define the net power-flux density as $j({\bf x},t)=\frac12\Re({\bf E}^*\times{\bf H})\cdot{\bf n}$,
with ${\bf n}={\bf k}/|{\bf k}|$ being the normal vector in the direction of the wave vector ${\bf k}$, 
the asterisk $*$ denoting complex conjugation
and $\Re$ denoting the real part. For transverse electric waves (row 1 in Table \ref{table1})
this reduces to $ j({\bf x},t)=\frac12\Re( E_y^* H_zn_x - E_y^* H_xn_z)$, 
and for transverse magnetic waves (row 2 in Table \ref{table1}) to $ j({\bf x},t)=\frac12\Re(- H_y^* E_zn_x+  H_y^* E_xn_z)$.
Both cases are captured by the compact definition
\begin{eqnarray}
 j({\bf x},t)=\frac12\Re( P^* {\bf Q})\cdot{\bf n}.\label{eqpowerq}
\end{eqnarray}
This equation also captures the net power-flux density 
for acoustic and horizontally polarized shear waves (rows 3 and 4 in Table \ref{table1}).

Next to the net power-flux density, we define the net field-momentum density for electromagnetic waves as 
$M({\bf x},t)=\frac12\Re({\bf D}^*\times{\bf B})\cdot{\bf n}$ \citep{Burns2020NJP, Feynmann63Book2}.
For transverse electric and transverse magnetic waves this can be written as
\begin{eqnarray}
 M({\bf x},t)=\frac12\Re( U^* {\bf V})\cdot{\bf n}.\label{eqmomentumq}
\end{eqnarray}
This equation also captures the net field-momentum density for acoustic and horizontally polarized shear waves 
(which should not be confused with the mechanical momentum density $m$ mentioned in Table \ref{table1}).

We analyse the quantities $j({\bf x},t)$ and $M({\bf x},t)$ for monochromatic plane waves before and after a time boundary (equations (\ref{eq3031}) -- (\ref{eq3036})).
Substituting $U=U_I$, ${\bf V}={\bf V}_I$ for $t<0$ and $U=U_T+U_R$, ${\bf V}={\bf V}_T+{\bf V}_R$ for $t>0$ 
 into equation (\ref{eqmomentumq}) we find, using equation (\ref{eq24u}), that  $M({\bf x},t)$ is constant for all ${\bf x}$ and $t$. 
Hence, the net field-momentum density is conserved when passing a time boundary.
Using equations (\ref{eq3}), (\ref{eq4}) and (\ref{eq8a}), it follows that $j({\bf x},t)$ and $M({\bf x},t)$ are mutually related via $j({\bf x},t)=c^2(t)M({\bf x},t)$.
Hence, $j_2=\frac{c_2^2}{c_1^2}j_1$, where $j_1$ and $j_2$ are the net power-flux densities before and after the time boundary.
Hence, the net power-flux density is not conserved when passing a time boundary.
This is  the result of energy being added to or extracted from the wave field by the mechanism that modulates the material parameters \citep{Morgenthaler58IRE, Mendonca2002PS, Caloz2020IEEE2}.

\section{Green's function}\label{sec4a}

We consider  a homogeneous, time-dependent material, with arbitrary piecewise continuous parameters $\alpha(t)$ and $\beta(t)$ (case iii in section \ref{sec2}).
We introduce the Green's function ${\cal G}({\bf x},{\bf x}_0,t,t_0)$ as the response to an impulsive point source at ${\bf x}={\bf x}_0$ at time instant $t=t_0$,
observed at position ${\bf x}$ and time $t$. Hence, it obeys wave
 equation (\ref{eq31a}), with the source term on the right-hand side replaced by an impulsive point source, according to
\begin{eqnarray}
\partial_t(\beta\partial_t{\cal G}) - \beta c^2{\bf \nabla}^2{\cal G} = \delta({\bf x}-{\bf x}_0)\delta(t-t_0),\label{eq31agbl}
\end{eqnarray}
where $\delta(t)$ is the Dirac delta function. Furthermore, the Green's function obeys the causality condition
\begin{eqnarray}
{\cal G}({\bf x},{\bf x}_0,t,t_0)=0 \quad \mbox{for}\quad t<t_0.\label{eq31con}
\end{eqnarray}
Assuming that the material is time-independent beyond an arbitrary large but finite time, this causality condition implies that ${\cal G}({\bf x},{\bf x}_0,t,t_0)$ is outward propagating for 
$|{\bf x}|\to\infty$. Since the material is homogeneous, the Green's function is shift-invariant, hence
\begin{eqnarray}
{\cal G}({\bf x},{\bf x}_0,t,t_0)={\cal G}({\bf x}-{\bf x}_0,{\bf 0},t,t_0).\label{eq31shf}
\end{eqnarray}

\begin{figure}
\centerline{\epsfxsize=15.5 cm \epsfbox{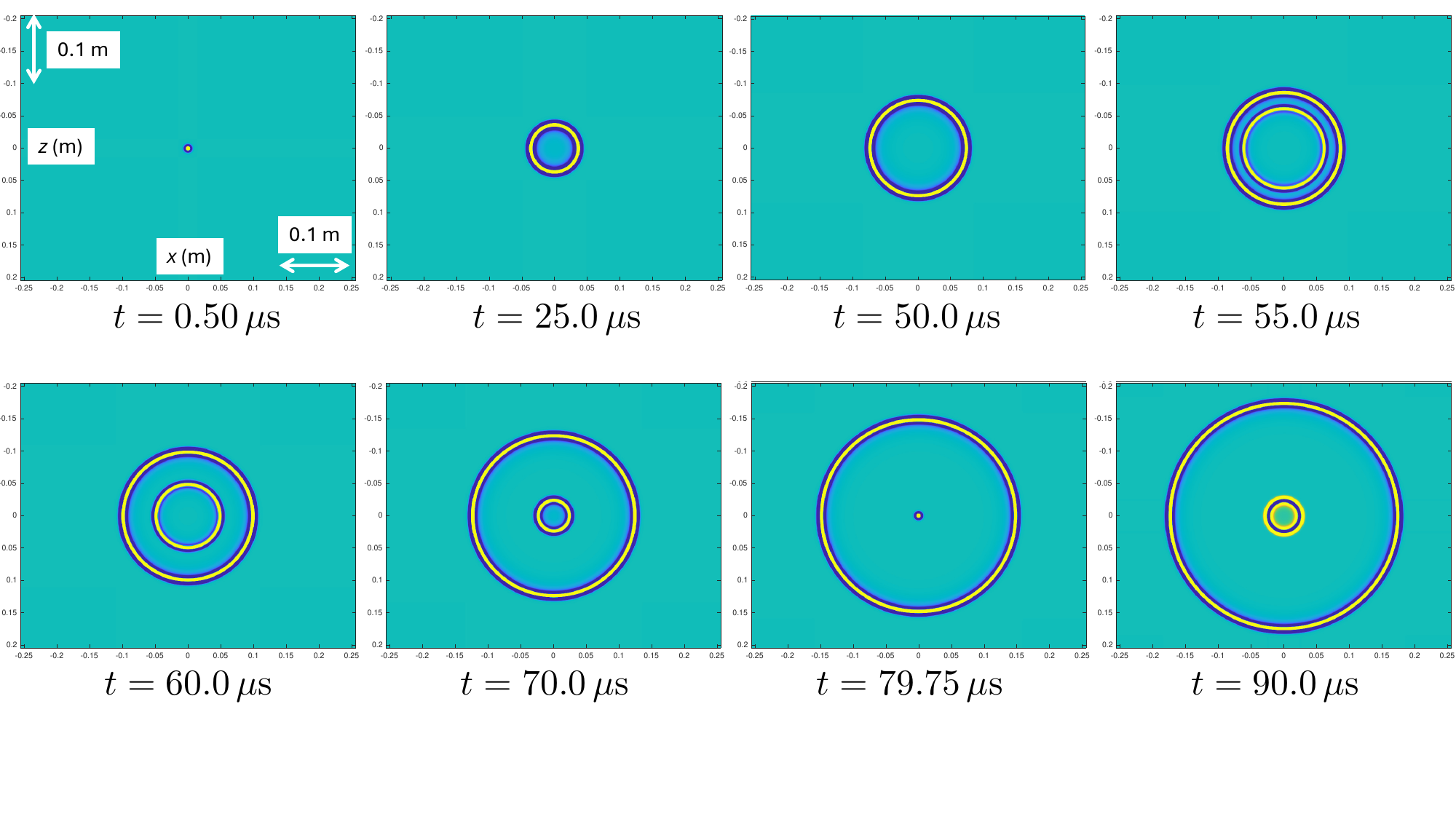}}
\vspace{-1cm}
\caption{Snapshots of the Green's function ${\cal G}({\bf x},{\bf 0},t,0)$ (convolved with a spatial wavelet) for a piecewise constant time-dependent material, with a time boundary at $t=50.0\,\mu$s
(see also movie Green.mp4 in the online material).}\label{Fig1}
\end{figure}

\begin{figure}
\centerline{\hspace{4.8cm}\epsfysize=7.5cm \epsfbox{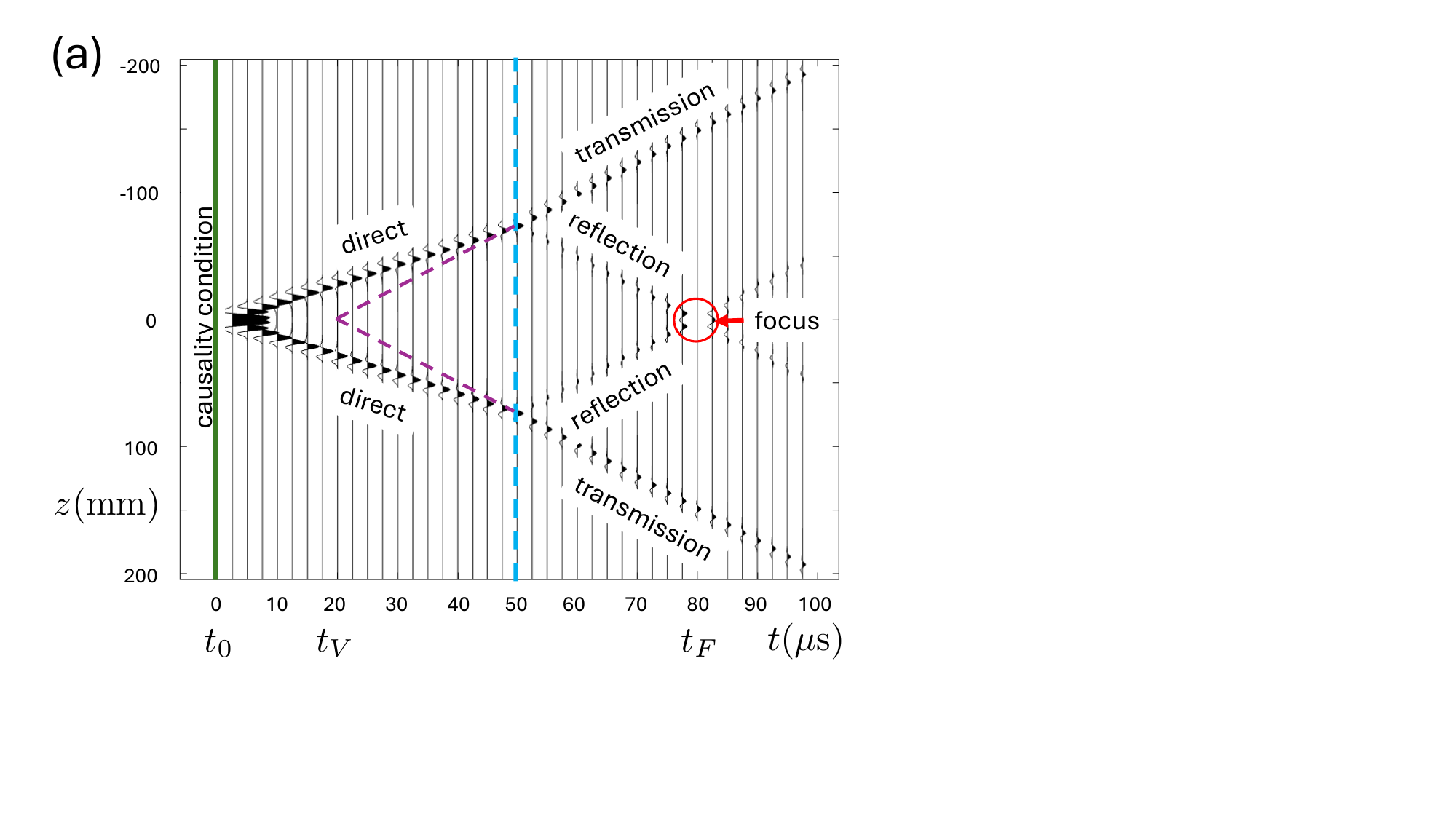}\,\hspace{-5.7cm}\epsfysize=7.5 cm \epsfbox{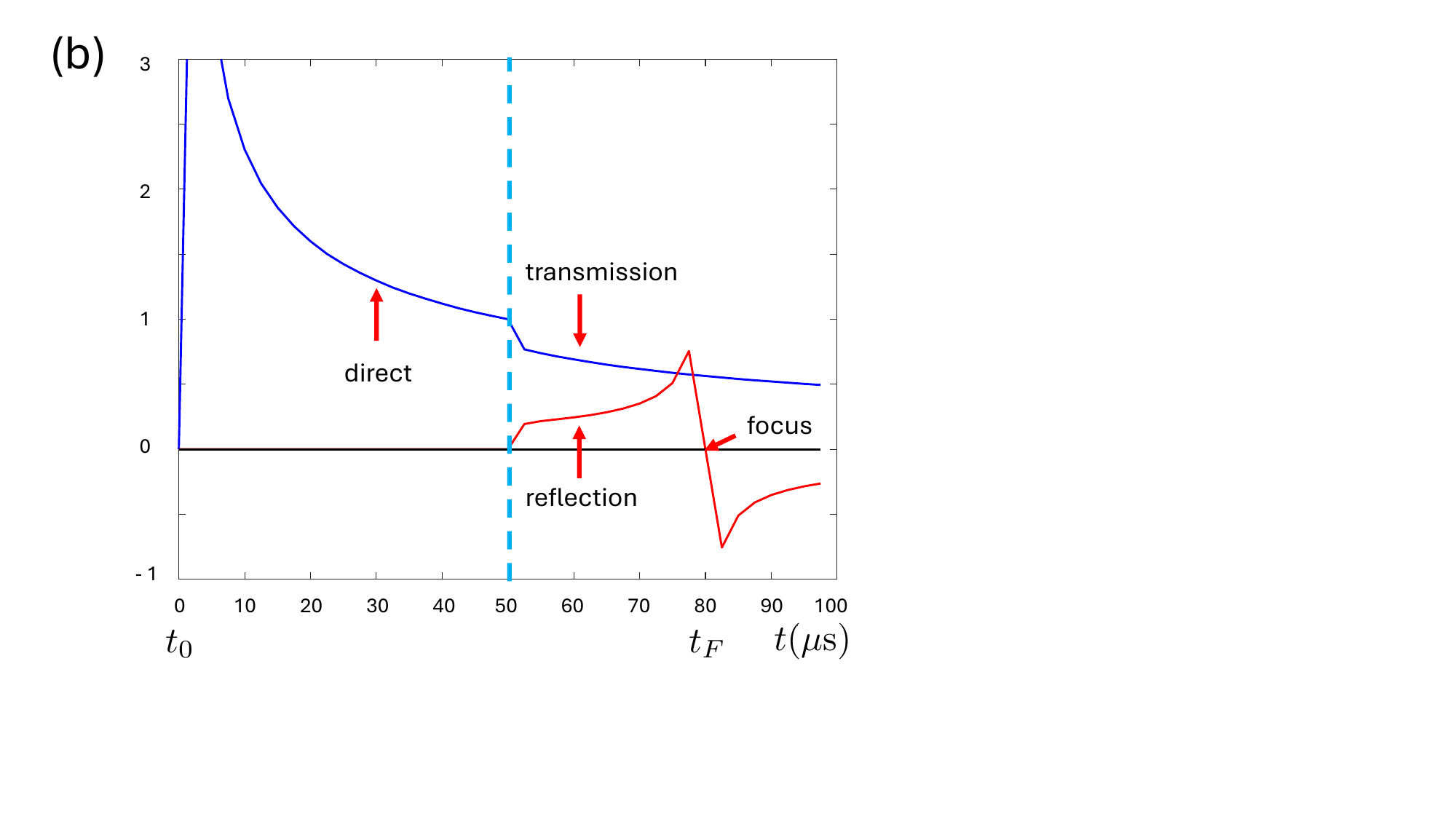}}
\vspace{-1.2cm}
\caption{(a) Cross-sections at $x=0$ of the Green's function of Figure \ref{Fig1}, i.e., ${\cal G}({\bf x},{\bf 0},t,0)|_{x=0}$.
(b) Scaled amplitude cross-sections of (a), measured along the direct and transmitted waves (blue) and along the reflected wave (red).}\label{Fig2}
\end{figure}

We discuss a numerical example of ${\cal G}({\bf x},{\bf x}_0,t,t_0)$ for a piecewise constant material with a single time boundary at $t=50\,\mu$s, 
with propagation velocities $c_1=1500$ m/s for $t<50\,\mu$s
and $c_2=2500$ m/s for $t>50\,\mu$s (the parameter $\beta$ is constant). 
The transmission and reflection coefficients at the time boundary, as defined in equations (\ref{eq90312}) and (\ref{eq90311}), are $T_u=0.8$ and $R_u=0.2$, respectively.
We choose the source at ${\bf x}_0={\bf 0}$ and $t_0=0$.
Figure \ref{Fig1} shows snapshots of ${\cal G}({\bf x},{\bf 0},t,0)$ (convolved with a 2D spatial wavelet with a central wavenumber $k_0/2\pi=100$ m$^{-1}$).
The first three frames show an expanding circular wave front, up to the time boundary at $t=50\,\mu$s. In the following frames the wave field is split into a transmitted wave front, 
which expands further, 
and a reflected wave front, which propagates back and collapses to a focus at ${\bf x}={\bf 0}$ and $t=79.75\,\mu$s. 
At exactly $t=80\,\mu$s the focused field is zero (not shown) and changes its sign, after which it
continues as an expanding circular wave front with opposite amplitude,
as shown in the last frame. Hence, the focus at ${\bf x}={\bf 0}$ and $t=80\,\mu$s acts as a virtual source for a wave field in a material with propagation velocity $c_2=2500$ m/s.
Figure \ref{Fig2}(a) shows cross-sections at $x=0$ of these snapshots in the form of a $z,t$-diagram 
(i.e., for ${\bf x}=(0,z)$, as if receivers were placed along a vertical line in the middle of these snapshots, through the source position). Left of the green line at $t_0=0$ the field is zero,
conform the causality condition of equation (\ref{eq31con}). Right of this green line, direct waves propagate away from the source, in two directions (up and down), 
until they reach the time boundary at $t=50\,\mu$s, indicated by the dashed blue line. Right of this blue line, transmitted and reflected waves can be seen.
Tracing the transmitted waves back in time along the purple dashed rays, they appear to originate from a virtual source at $z=0$ and $t_V=20\,\mu$s.
The reflected waves propagate back and focus at the position of the original source at $z=0$ and $t_F=80\,\mu$s. 
This focus acts as a virtual source for the reflected waves beyond $t_F=80\,\mu$s.
Figure \ref{Fig2}(b) shows an amplitude cross-section of Figure \ref{Fig2}(a), scaled such that the amplitude of the direct wave arriving at the time boundary at $t=50\,\mu$s equals 1.
The blue curve shows that the amplitudes of the direct and transmitted waves gradually decay with time, and jump from 1 to $T_u=0.8$ across the time  boundary.  
The red curve shows that the reflected wave starts directly after the time boundary with an amplitude of $R_u=0.2$. 
Its amplitude initially increases during back-propagation, changes sign at the focus and  decreases after the focus.

Whereas the Green's function ${\cal G}({\bf x},{\bf x}_0,t,t_0)$ is the response to an impulsive point source $\delta({\bf x}-{\bf x}_0)\delta(t-t_0)$ 
for arbitrary ${\bf x}_0$ and $t_0$ (equation (\ref{eq31agbl})), the wave field $U({\bf x},t)$ is the response to a source distribution $s({\bf x},t)$ (equation (\ref{eq31a})).
We obtain a simple representation for the wave field $U$ in terms of ${\cal G}$, assuming they both reside in the same medium and both are outward propagating for $|{\bf x}|\to\infty$.
Since both equations are linear, we obtain the representation for $U({\bf x},t)$ by applying Huygens' superposition principle, according to \citep{Morse53Book, Bleistein84Book}
\begin{eqnarray}
U({\bf x},t)&=& \int_{-\infty}^\infty{\rm d}t_0 \int_{{\mathbb{R}}^2}{\cal G}({\bf x},{\bf x}_0,t,t_0)s({\bf x}_0,t_0){\rm d}{\bf x}_0,\label{eq31kbl}
\end{eqnarray}
where  ${\bf x}_0$ and $t_0$ are variables and 
where ${\mathbb R}$ is the set of real numbers.
This is a special case of the more general representation derived in section \ref{sec7.2}.
Using equation (\ref{eq31shf}), we can rewrite equation (\ref{eq31kbl}) as
\begin{eqnarray}
U({\bf x},t)&=&\int_{-\infty}^\infty{\rm d}t_0 \int_{{\mathbb{R}}^2}{\cal G}({\bf x}-{\bf x}_0,{\bf 0},t,t_0)s({\bf x}_0,t_0){\rm d}{\bf x}_0.\label{eq31kblconv}
\end{eqnarray}
Note that the space integral represents a 2D space convolution. Since the material is time-dependent, the time integral cannot be rewritten as a genuine time convolution.

\begin{figure}
\centerline{\epsfxsize=15.5 cm \epsfbox{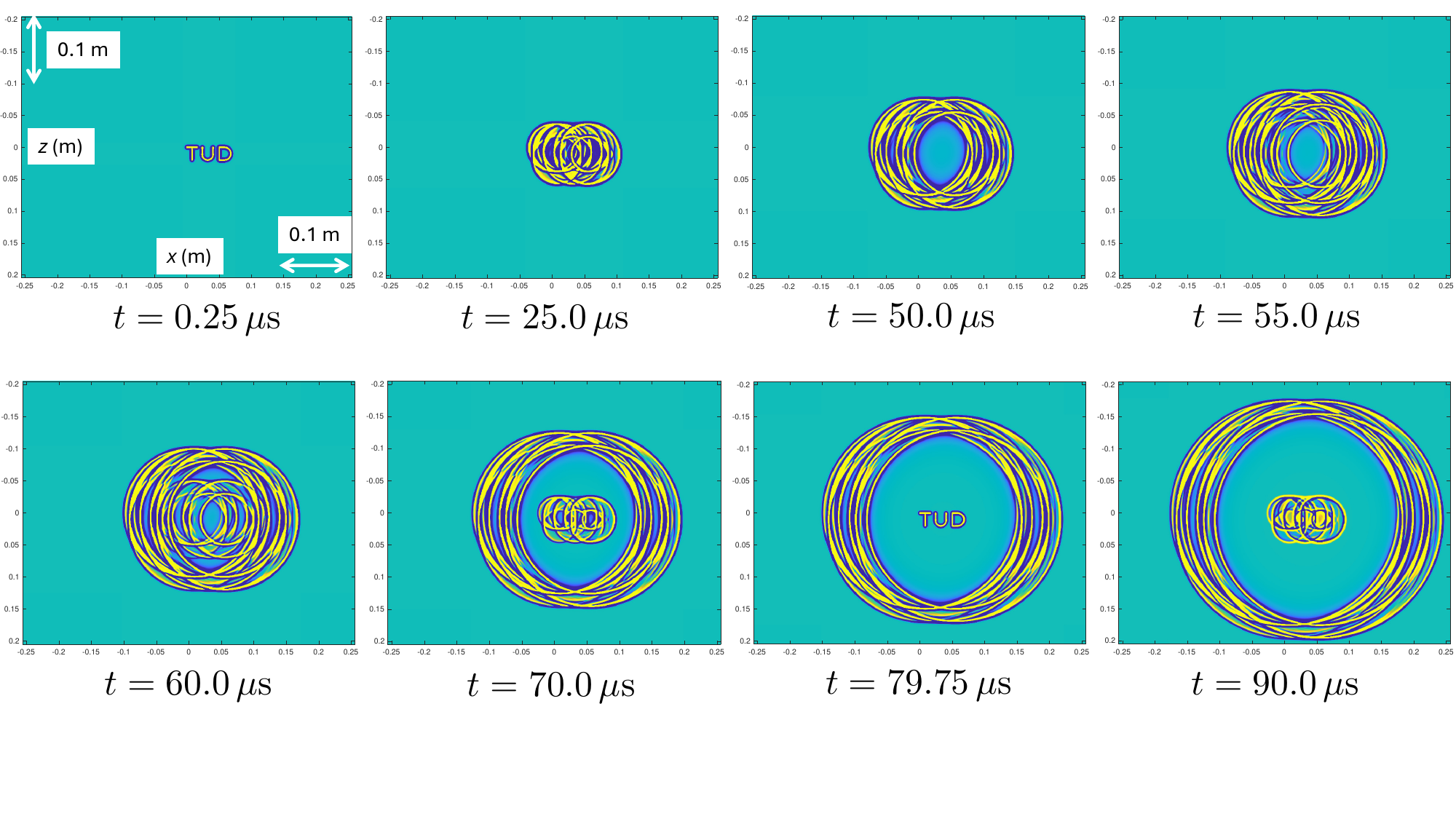}}
\vspace{-1cm}
\caption{The wave field $U({\bf x},t)$ in response to a distributed source in the same piecewise constant time-dependent material as in Figure \ref{Fig1}
(see also movie TUD.mp4 in the online material).}\label{Fig3}
\end{figure}

Figure \ref{Fig3} is an illustration of equation (\ref{eq31kblconv}) for a source distribution 
$s({\bf x},t)=s_0({\bf x})\delta(t)$, i.e., for a distributed source $s_0$ as a function of ${\bf x}$ at $t=0$
(this time the 2D spatial wavelet has a central wavenumber $k_0/2\pi=200$ m$^{-1}$).
The first frame shows the response at $t=0.25\,\mu$s to this distributed source in the same time-dependent material as in the previous example. The next frames show the evolution
of this response over time. After the time boundary at $t=50\,\mu$s, the transmitted field continues expanding, whereas the reflected field propagates back and 
collapses to a reproduction of the distributed source at $t=79.75\,\mu$s. After this, the focused field changes its sign (at $t=80\,\mu$s) and expands again. 
Results like this have been obtained with physical experiments by  \cite{Bacot2016NP, Fink2017EPJ, Peng2020JPC}.

Finally, for the special case of a time-independent medium, the analytical solution of equation (\ref{eq31agbl}), with causality condition (\ref{eq31con}), reads
\begin{eqnarray}
{\cal G}({\bf x},{\bf x}_0,t,t_0)=\frac{1}{2\pi\beta c^2}\frac{H\Bigl(t-t_0- \frac{|{\bf x}-{\bf x}_0|}{c}\Bigr)}{\sqrt{(t-t_0)^2- \frac{|{\bf x}-{\bf x}_0|^2}{c^2}}},\label{eq71bl}
\end{eqnarray}
where $H(t)$ is the Heaviside step function. The direct waves in Figure \ref{Fig2}(a) can be seen as an illustration of this expression for ${\cal G}({\bf x},{\bf 0},t,0)|_{x=0}$, convolved with a spatial wavelet.

\section{Unified wave equation and Green's function in wave-vector time domain}

\subsection{Unified wave equation in wave-vector time domain}
We define the following 2D Fourier transformation
\begin{eqnarray}
\check U({\bf k},t) =\int_{{\mathbb{R}}^2} U({\bf x},t)\exp (-i{\bf k}\cdot{\bf x}){\rm d}{\bf x},\label{eq51}
\end{eqnarray}
with wave vector ${\bf k}$ defined as ${\bf k}=(k_x,k_z)$. For the inverse 2D Fourier transformation we  obtain
\begin{eqnarray}
U({\bf x},t)=\frac{1}{4\pi^2} \int_{{\mathbb{R}}^2} \check U({\bf k},t)  \exp(i{\bf k}\cdot{\bf x}){\rm d}{\bf k}. \label{eq52}
\end{eqnarray}
Equation (\ref{eq52}) expresses $U({\bf x},t)$ as a superposition of ``space-harmonic'' plane waves $\exp(i{\bf k}\cdot{\bf x})$ with complex amplitudes $\check U({\bf k},t)$.
Hence, equation (\ref{eq51}) can be interpreted as a decomposition of the field $U({\bf x},t)$ into space-harmonic plane waves.

Applying the nabla operator ${\bf \nabla}$ to both sides of equation (\ref{eq52}) yields 
\begin{eqnarray}
{\bf \nabla}U({\bf x},t)=\frac{1}{4\pi^2} \int_{{\mathbb{R}}^2} i{\bf k}\check U({\bf k},t)  \exp(i{\bf k}\cdot{\bf x}){\rm d}{\bf k}. \label{eq59}
\end{eqnarray}
Hence, the operation ${\bf \nabla}$ in the space-time domain corresponds to multiplication by $i{\bf k}$ in the wave-vector time domain. Using this property, we obtain 
 for the Fourier transforms of equations (\ref{eq49}) and (\ref{eq50}) 
\begin{eqnarray}
\partial_t \check U + \frac{1}{\beta}i{\bf k}\cdot \check {\bf V} &=& \check a,\label{eq53}\\
\partial_t \check {\bf V} + \frac{1}{\alpha}i{\bf k}\check U &=& \check {\bf b},\label{eq54}
\end{eqnarray}
with time-dependent material parameters $\alpha(t)$ and $\beta(t)$. 
When the material contains time boundaries, the boundary conditions state that  $\check U({\bf k},t)$ and $\check {\bf V}({\bf k},t)$ are continuous over those time boundaries.
The second order wave equation (\ref{eq31a}) transforms to
\begin{eqnarray}
\partial_t(\beta\partial_t\check U) + \beta c^2|{\bf k}|^2\check U =\check s,\label{eq156bl}
\end{eqnarray}
with $|{\bf k}|^2={\bf k}\cdot{\bf k}=k_x^2+k_z^2$, and the source function $\check s({\bf k},t)$ defined as
\begin{eqnarray}
\check s= \partial_t(\beta\check a) - i{\bf k}\cdot\check {\bf b}.\label{eq2156bl}
\end{eqnarray}

\subsection{Green's function in wave-vector time domain}

The transformed Green's function $\check {\cal G}({\bf k},{\bf x}_0,t,t_0)$ obeys
\begin{eqnarray}
\partial_t(\beta\partial_t\check {\cal G}) +\beta c^2|{\bf k}|^2\check {\cal G} =  \exp(-i{\bf k}\cdot{\bf x}_0)\delta(t-t_0),\label{eq3156bl}
\end{eqnarray}
with causality condition 
\begin{eqnarray}
\check {\cal G}({\bf k},{\bf x}_0,t,t_0)=0\quad \mbox{for}\quad t<t_0.\label{eqcaus2bl}
\end{eqnarray}
For the special case of a time-independent medium we have
\begin{eqnarray}
\check {\cal G}({\bf k},{\bf x}_0,t,t_0)=\exp(-i{\bf k}\cdot{\bf x}_0)H(t-t_0)\frac{\sin(|{\bf k}|c(t-t_0))}{\eta|{\bf k}|}.\label{eq93bebl}
\end{eqnarray}
In Appendix \ref{App2} it is shown that the inverse Fourier transform of equation (\ref{eq93bebl}) is the space-time domain Green's function of equation (\ref{eq71bl}).

\section{Propagation invariants for piecewise continuous materials}\label{sec6}

\subsection{Conservation of net field-momentum density}

Analogous to equations (\ref{eqpowerq}) and (\ref{eqmomentumq}), we define the net power-flux density and the net field-momentum density 
in the ${\bf k},t$-domain as 
\begin{eqnarray}
\check \jmath({\bf k},t)=\frac12\Re(\check P^*\check {\bf Q})\cdot{\bf n}\label{eqpower}
\end{eqnarray}
and
\begin{eqnarray}
\check M({\bf k},t)=\frac12\Re(\check U^*\check {\bf V})\cdot{\bf n},\label{eqmomentum}
\end{eqnarray}
respectively, with ${\bf n}={\bf k}/|{\bf k}|$ (but note that $\check \jmath({\bf k},t)$ and $\check M({\bf k},t)$ are not the spatial Fourier transforms of $j({\bf x},t)$ and $M({\bf x},t)$).

Propagation invariants in a time-dependent material are wave-field-related quantities that remain constant over time.
We start by showing that for a source-free, time-dependent material, the net field-momentum density defined in equation (\ref{eqmomentum}) is a propagation invariant.
To this end we analyze $\partial_t\check M({\bf k},t)=\frac14\partial_t(\check U^*\check{\bf V}+\check{\bf V}^*\check U)\cdot{\bf n}$. Applying the product rule for differentiation,
using equations (\ref{eq53}) and (\ref{eq54}) for the source-free situation and the definition ${\bf n}={\bf k}/|{\bf k}|$, we obtain
\begin{eqnarray}
\partial_t\check M({\bf k},t)
&=&\frac14\Bigl((\partial_t\check U)^*\check{\bf V}+\check U^*\partial_t\check{\bf V}
+(\partial_t\check {\bf V})^*\check U+\check {\bf V}^*\partial_t\check U\Bigr)\cdot{\bf k}/|{\bf k}|\label{eq122bl}\\
&=&\frac14\biggl(\Bigl(\frac{1}{\beta}i{\bf k}\cdot\check {\bf V}^*\Bigr)\check{\bf V}-\check U^*\Bigl(\frac{1}{\alpha}i{\bf k}\check U\Bigr)
+\Bigl(\frac{1}{\alpha}i{\bf k}\check U^*\Bigr)\check U-\check {\bf V}^*\Bigl(\frac{1}{\beta}i{\bf k}\cdot\check {\bf V}\Bigr)\biggr)\cdot{\bf k}/|{\bf k}|=0.\nonumber
\end{eqnarray}
This expression implies that $\check M({\bf k},t)$ is a propagation invariant, or in other words, the net field-momentum density is a conserved quantity. 
The derivation above holds for continuously varying material parameters. 
However,  $\check U$ and $\check{\bf V}\cdot{\bf n}$ are continuous over a time boundary and so is $\check M({\bf k},t)$. 
Combining these results, it follows that $\check M({\bf k},t)$ is a propagation invariant for 
a time-dependent material with piecewise continuous parameters.

From equations (\ref{eq3}), (\ref{eq4}), (\ref{eq8a}), (\ref{eqpower}) and (\ref{eqmomentum}) we obtain
\begin{eqnarray}
\check \jmath({\bf k},t)=c^2(t)\check M({\bf k},t),
\end{eqnarray}
which implies that for a time-dependent medium with piecewise continuous parameters 
the net power-flux density is not a propagation invariant, similar as  noted in section \ref{sec3b} for $j({\bf x},t)$ over a single time boundary.
 \cite{Torrent2018PRB} shows that amplitude amplification due to temporal variations can be used to compensate for energy loss due to material dissipation.

\subsection{General propagation invariants}
Next, we derive more general propagation invariants. We consider two mutually independent states (i.e., two 
solutions of equations (\ref{eq53}) and (\ref{eq54})), which we distinguish with subscripts $A$ and $B$.
We analyze the quantities $\partial_t(\check U_A\check {\bf V}_B - \check {\bf V}_A\check U_B)\cdot(i{\bf k})$ and $\partial_t(\check U_A^*\check {\bf V}_B + \check {\bf V}_A^*\check U_B)\cdot(i{\bf k})$.
The latter can be seen as a generalization of $4\partial_t\check M({\bf k},t)=\partial_t(\check U^*\check{\bf V}+\check{\bf V}^*\check U)\cdot{\bf n}$, 
with subscripts $A$ and $B$ added and  ${\bf n}={\bf k}/|{\bf k}|$ replaced by $i{\bf k}$ (the latter replacement facilitates the inverse Fourier transformations in later sections).
Applying the product rule for differentiation, using equations (\ref{eq53}) and (\ref{eq54}) for states $A$ and $B$, yields 
\begin{eqnarray}
\partial_t(\check U_A\check {\bf V}_B - \check {\bf V}_A\check U_B)\cdot(i{\bf k}) &=&
|{\bf k}|^2(\alpha_B^{-1}-\alpha_A^{-1})\check U_A\check U_B
+(\beta_B^{-1}-\beta_A^{-1})\check V_{{\rm d},A}\check V_{{\rm d},B}\nonumber\\
&&+\check a_A\check V_{{\rm d},B}-\check b_{{\rm d},A}\check U_B-\check V_{{\rm d},A}\check a_B+\check U_A\check b_{{\rm d},B},\label{eq1013tbl}
\end{eqnarray}
and
\begin{eqnarray}
\partial_t(\check U_A^*\check {\bf V}_B + \check {\bf V}_A^*\check U_B)\cdot(i{\bf k})&=&
|{\bf k}|^2(\alpha_B^{-1}-\alpha_A^{-1})\check U_A^*\check U_B
+(\beta_B^{-1}-\beta_A^{-1})\check V_{{\rm d},A}^*\check V_{{\rm d},B}\nonumber\\
&&+\check a_A^*\check V_{{\rm d},B}-\check b_{{\rm d},A}^*\check U_B-\check V_{{\rm d},A}^*\check a_B+\check U_A^*\check b_{{\rm d},B},\label{eq1014tbl}
\end{eqnarray}
respectively, with
\begin{eqnarray}
\check V_{\rm d}&=&i{\bf k}\cdot\check {\bf V},\label{eq457}\\
\check b_{\rm d}&=&i{\bf k}\cdot\check {\bf b},\label{eq458}
\end{eqnarray}
for states $A$ and $B$.
The subscript ${\rm d}$ in $\check V_{\rm d}$ and $\check b_{\rm d}$ refers to ``divergence'' 
(since $i{\bf k}\cdot\check {\bf V}$ and $i{\bf k}\cdot\check {\bf b}$ are the Fourier transforms of ${\nabla}\cdot{\bf V}$ and ${\nabla}\cdot{\bf b}$).
Following similar arguments as below equation (\ref{eq122bl}), equations (\ref{eq1013tbl}) and (\ref{eq1014tbl})
hold for a time-dependent material with piecewise continuous parameters.

Note that in equations (\ref{eq1013tbl}) and (\ref{eq1014tbl}) not only the wave fields are labeled with subscripts $A$ and $B$, 
but also the source terms and the material parameters (which may be different in the two states).
In the next section we continue with the general expressions of equations (\ref{eq1013tbl}) and (\ref{eq1014tbl}). 
Here we consider the special case that sources are absent and the material parameters are the same in both states, 
which implies that the right-hand sides of equations (\ref{eq1013tbl}) and (\ref{eq1014tbl}) vanish.
Hence, for this situation it follows that $(\check U_A\check {\bf V}_B - \check {\bf V}_A\check U_B)\cdot(i{\bf k})$ and $(\check U_A^*\check {\bf V}_B + \check {\bf V}_A^*\check U_B)\cdot(i{\bf k})$ are 
propagation invariants for a homogeneous time-dependent material. They are the counterparts of propagation invariants for an inhomogeneous time-independent material
 \citep{Haines88GJI, Kennett90GJI, Koketsu91GJI, Takenaka93WM, Wapenaar2004GJI}, 
which are useful for the analysis of symmetry properties of transmission and reflection responses and for the design of efficient numerical modelling schemes.

\section{Reciprocity theorems}

\subsection{General reciprocity theorems}

Reciprocity theorems interrelate wave fields in two different states \citep{Rayleigh78Book, Lorentz1895KNAW, Knopoff59GEO, Hoop66ASR}. Here we use equations (\ref{eq1013tbl}) and (\ref{eq1014tbl}) to derive
 reciprocity theorems for a homogeneous time-dependent material  with piecewise continuous parameters $\alpha(t)$ and $\beta(t)$.
Integrating equations (\ref{eq1013tbl}) and (\ref{eq1014tbl}) with respect to time, 
taking into account that $\check U$ and $\check V_{\rm d}=i{\bf k}\cdot\check {\bf V}$ are continuous at time instants where $\alpha(t)$ and $\beta(t)$ are discontinuous, we obtain
\begin{eqnarray}
(\check U_A\check V_{{\rm d},B} - \check V_{{\rm d},A}\check U_B)\Bigr|_{t_b}^{t_e} &=&\int_{t_b}^{t_e}\Bigl(|{\bf k}|^2(\alpha_B^{-1}-\alpha_A^{-1})\check U_A\check U_B
+(\beta_B^{-1}-\beta_A^{-1})\check V_{{\rm d},A}\check V_{{\rm d},B}\nonumber\\
&&+\check a_A\check V_{{\rm d},B}-\check b_{{\rm d},A}\check U_B-\check V_{{\rm d},A}\check a_B+\check U_A\check b_{{\rm d},B}\Bigr){\rm d}t\label{eq1017tbl}
\end{eqnarray}
and
\begin{eqnarray}
(\check U_A^*\check V_{{\rm d},B} - \check V_{{\rm d},A}^*\check U_B)\Bigr|_{t_b}^{t_e} &=&\int_{t_b}^{t_e}\Bigl(|{\bf k}|^2(\alpha_B^{-1}-\alpha_A^{-1})\check U_A^*\check U_B
+(\beta_B^{-1}-\beta_A^{-1})\check V_{{\rm d},A}^*\check V_{{\rm d},B}\nonumber\\
&&+\check a_A^*\check V_{{\rm d},B}-\check b_{{\rm d},A}^*\check U_B-\check V_{{\rm d},A}^*\check a_B+\check U_A^*\check b_{{\rm d},B}\Bigr){\rm d}t,\label{eq1017tcbl}
\end{eqnarray}
respectively, with subscripts $b$ and $e$ standing for ``begin'' and ``end''. Products like $\check U_A\check U_B$ in the ${\bf k},t$-domain correspond to spatial convolutions
in the ${\bf x},t$-domain. Therefore we call equation (\ref{eq1017tbl}) the reciprocity theorem of the space-convolution type. It is the counterpart of the reciprocity theorem of the 
time-convolution type for an inhomogeneous time-independent medium \citep{Hoop95Book, Fokkema93Book}.
On the other hand, products like $\check U_A^*\check U_B$ in the ${\bf k},t$-domain correspond to spatial correlations
in the ${\bf x},t$-domain. Therefore we call equation (\ref{eq1017tcbl}) the reciprocity theorem of the space-correlation type. It is the counterpart of the reciprocity theorem of the 
time-correlation type for an inhomogeneous time-independent medium \citep{Hoop95Book, Bojarski83JASA}.

We use equation (\ref{eq1017tbl}) in section \ref{sec7.2} to derive a general wave field representation and in section \ref{sec8}  we use equation (\ref{eq1017tcbl})
to derive an expression for Green's function retrieval, both for time-dependent materials. Here we discuss two special cases.

\subsection{Source-receiver reciprocity}

First, we derive from equation (\ref{eq1017tbl}) a reciprocity relation for the Green's function $\check {\cal G}({\bf k},{\bf x}_A,t,t_A)$.
Table \ref{table2} lists the quantities for states $A$ and $B$ that we use in this derivation.
The material parameters are chosen the same in both states. 
For the wave field $\check U_A$ we choose the acausal Green's function $\check {\cal G}^a({\bf k},{\bf x}_A,t,t_A)$. 
It obeys the same wave equation as the causal Green's function (equation (\ref{eq3156bl})), but with the condition $\check {\cal G}^a({\bf k},{\bf x}_A,t,t_A)=0$ for $t>t_A$.
For the wave field $\check U_B$ we choose the causal Green's function $\check {\cal G}({\bf k},{\bf x}_B,t,t_B)$, with $\check {\cal G}({\bf k},{\bf x}_B,t,t_B)=0$ for $t<t_B$.
The expressions in Table \ref{table2} for the wave fields $\check V_{{\rm d},A}$ and $\check V_{{\rm d},B}$ follow from equations (\ref{eq53}) and (\ref{eq457}) for states $A$ and $B$ 
(using $\check a_A=\check a_B=0$). The expressions for the source terms $\check b_{{\rm d},A}$ and $\check b_{{\rm d},B}$ 
 follow from comparing the source term for a Green's function (equation (\ref{eq3156bl})) with that for a general wave field (equations (\ref{eq2156bl}) and (\ref{eq458}),
 using $\check a_A=\check a_B=0$). Note that the source term for the acausal Green's function in state $A$ actually represents a sink.
 
\begin{table}
\caption{
Quantities for deriving the reciprocity relation of equation (\ref{eq1022taqbl}).}\label{table2}
\begin{tabular}
{lll}
\hline
& {\rm State $A$} & {\rm State $B$} \\
\hline
Material parameters & $\alpha_A=\alpha_A(t)$ & $\alpha_B=\alpha_A(t)$  \\
 & $\beta_A=\beta_A(t)$ & $\beta_B=\beta_A(t)$  \\
Wave fields & $\check U_A=\check {\cal G}^a({\bf k},{\bf x}_A,t,t_A)$ & $\check U_B=\check {\cal G}({\bf k},{\bf x}_B,t,t_B)$\\
& $\check V_{{\rm d},A}=-\beta_A(t)\partial_t\check {\cal G}^a({\bf k},{\bf x}_A,t,t_A)$ & $\check V_{{\rm d},B}=-\beta_A(t)\partial_t\check {\cal G}({\bf k},{\bf x}_B,t,t_B)$\\
Sources & $\check a_A=0$ & $\check a_B=0$ \\
 & $\check b_{{\rm d},A}=-\exp(-i{\bf k}\cdot{\bf x}_A)\delta(t-t_A)$ & $\check b_{{\rm d},B}=-\exp(-i{\bf k}\cdot{\bf x}_B)\delta(t-t_B)$ \\
\hline
\end{tabular}
\end{table}

We substitute the quantities of Table \ref{table2} into the reciprocity theorem of the space-convolution type, formulated by equation (\ref{eq1017tbl}).
 Let us assume that the time instants $t_A$ and $t_B$ lie both between  $t_b$ and $t_e$.
Then the acausal Green's function is zero at $t_e$ and the causal Green's function is zero at $t_b$. With this, the left-hand side of equation (\ref{eq1017tbl}) vanishes. 
From the remaining terms on the right-hand side of equation (\ref{eq1017tbl}) we obtain
\begin{eqnarray}
\exp(i{\bf k}\cdot{\bf x}_A)\check {\cal G}^a({\bf k},{\bf x}_A,t_B,t_A)=\exp(i{\bf k}\cdot{\bf x}_B)\check {\cal G}({\bf k},{\bf x}_B,t_A,t_B).\label{eq1022taqbl}
\end{eqnarray}
This is the sought reciprocity relation for the Green's function in the ${\bf k},t$-domain.
We use the inverse Fourier transformation defined in equation (\ref{eq52}) to transform this expression to the ${\bf x},t$-domain. This yields
\begin{eqnarray}
{\cal G}^a({\bf x}+{\bf x}_A,{\bf x}_A,t_B,t_A)={\cal G}({\bf x}+{\bf x}_B,{\bf x}_B,t_A,t_B),\label{eq1022sqbl}
\end{eqnarray}
or, since in a homogeneous medium the Green's function is shift invariant,
\begin{eqnarray}
{\cal G}^a({\bf x},{\bf 0},t_B,t_A)={\cal G}({\bf x},{\bf 0},t_A,t_B).\label{eq1022sqshbl}
\end{eqnarray}
This is the counterpart of the well-known source-receiver reciprocity relation ${\cal G}({\bf x}_B,{\bf x}_A,t,0)={\cal G}({\bf x}_A,{\bf x}_B,t,0)$ 
for an inhomogeneous time-independent medium, which states that 
the Green's function between a source at ${\bf x}_B$ and a receiver at ${\bf x}_A$ is identical to the Green's function between a source at ${\bf x}_A$ and a receiver at ${\bf x}_B$
\citep{Hoop95Book, Morse53Book}. 
Equation  (\ref{eq1022sqshbl}) states that for a homogeneous time-dependent medium 
the causal Green's function between a source at $t_B$ and a receiver at $t_A$ is identical to the {\it acausal} Green's function between a sink at $t_A$ and a receiver at $t_B$.
In the derivation we assumed that $t_A$ and $t_B$ lie both between  $t_b$ and $t_e$. If we take $t_b\to-\infty$ and $t_e\to\infty$, then
equations (\ref{eq1022taqbl}) -- (\ref{eq1022sqshbl}) hold for arbitrary $t_A$ and $t_B$. Note that for $t_A<t_B$ these expressions reduce to the trivial relation $0=0$. 

\subsection{Field-momentum density balance}

The next special case we consider is the field-momentum density balance, which we derive from equation (\ref{eq1017tcbl}).
We take identical states $A$ and $B$ and drop the subscripts. We thus obtain from equation (\ref{eq1017tcbl})
\begin{eqnarray}
(\check U^*\check V_{\rm d}-\check V_{\rm d}^*\check U)\Bigr|_{t_b}^{t_e}=
\int_{t_b}^{t_e}\bigl(\check a^*\check V_{\rm d}-\check b_{\rm d}^*\check U-\check V_{\rm d}^*\check a+\check U^*\check b_{\rm d}\bigr){\rm d}t,
\end{eqnarray}
or, using equations (\ref{eqmomentum}) and (\ref{eq457}) and the definition ${\bf n}={\bf k}/|{\bf k}|$,
\begin{eqnarray}
\check M({\bf k},t)\Bigr|_{t_b}^{t_e} &=&\frac12\Re\int_{t_b}^{t_e}\bigl(\check a^*\check {\bf V}+\check {\bf b}^*\check U\bigr)\cdot{\bf n}{\rm d}t.\label{eq1017tdbl}
\end{eqnarray}
This is the net field-momentum density balance for a homogeneous time-dependent material for the situation that there are sources between $t_b$ and $t_e$.
It is the counterpart of the net powerflux density balance for an inhomogeneous time-independent medium \citep{Hoop95Book, Fokkema93Book}.

\section{Wave field representations}\label{sec7.2}

\subsection{General wave field representation}

A wave field representation is obtained by replacing one of the states in a reciprocity theorem by a Green's state \citep{Knopoff56JASA, Hoop58PHD, Gangi70JGR, Pao76JASA}.
We derive a general wave field representation for a homogeneous time-dependent material  with piecewise continuous parameters $\alpha(t)$ and $\beta(t)$.
To this end, consider the quantities listed in Table \ref{table3}.
Note that state $A$ is the same as that in Table \ref{table2}, except that ${\bf x}_A$ and $t_A$ are for convenience replaced by ${\bf 0}$ and $t'$, respectively.
State $B$ is the actual field in the actual medium, so the subscripts $B$ are dropped. 

\begin{table}
\caption{
Quantities for deriving the representation of equation (\ref{eq65kbl}).}\label{table3}
\begin{tabular}
{lll}
\hline
& {\rm State $A$} & {\rm State $B$} \\
\hline
Material parameters & $\alpha_A=\alpha_A(t)$ & $\alpha_B=\alpha(t)$  \\
 & $\beta_A=\beta_A(t)$ & $\beta_B=\beta(t)$  \\
Wave fields & $\check U_A=\check {\cal G}^a({\bf k},{\bf 0},t,t')$ & $\check U_B=\check U({\bf k},t)$\\
& $\check V_{{\rm d},A}=-\beta_A(t)\partial_t\check {\cal G}^a({\bf k},{\bf 0},t,t')$ & $\check V_{{\rm d},B}=\check V_{\rm d}({\bf k},t)$\\
Sources & $\check a_A=0$ & $\check a_B=\check a({\bf k},t)$ \\
 & $\check b_{{\rm d},A}=-\delta(t-t')$ & $\check b_{{\rm d},B}=\check b_{\rm d}({\bf k},t)$ \\
\hline
\end{tabular}
\end{table}

Substitution of the quantities in Table \ref{table3} into the reciprocity theorem of the space-convolution type, formulated by equation (\ref{eq1017tbl}), 
using reciprocity relation (\ref{eq1022taqbl}) for the Green's function with ${\bf x}_A={\bf x}_B={\bf 0}$, yields
\begin{eqnarray}
&&\hspace{-.7cm}\chi(t')\check U({\bf k},t')=
\Bigl(\check {\cal G}({\bf k},{\bf 0},t',t)\check V_{\rm d}({\bf k},t)+\beta_A(t)\{\partial_t\check {\cal G}({\bf k},{\bf 0},t',t)\}\check U({\bf k},t)\Bigr)\Bigr|_{t=t_b}^{t=t_e}\nonumber\\
&&-\int_{t_b}^{t_e}\Bigl(|{\bf k}|^2\check {\cal G}({\bf k},{\bf 0},t',t)\Delta\alpha^{-1}(t)\check U({\bf k},t)-
\beta_A(t)\{\partial_t\check {\cal G}({\bf k},{\bf 0},t',t)\}\Delta\beta^{-1}(t)\check V_{\rm d}({\bf k},t)\Bigr){\rm d}t\nonumber\\
&&-\int_{t_b}^{t_e}\Bigl(\{\partial_t\check {\cal G}({\bf k},{\bf 0},t',t)\}\beta_A(t)\check a({\bf k},t)+\check {\cal G}({\bf k},{\bf 0},t',t)\check b_{\rm d}({\bf k},t)\Bigr){\rm d}t,\label{eq65kbl}
\end{eqnarray}
with the characteristic function $\chi(t')$ defined as
\begin{eqnarray}
\chi(t')=\begin{cases}
&1\quad\mbox{for}\quad t_b<t'<t_e,\\
&\frac12 \quad\mbox{for}\quad t'=t_b \quad\mbox{or}\quad t'=t_e,\\
&0 \quad\mbox{for}\quad t'<t_b \quad\mbox{or}\quad t'>t_e,
\end{cases}
\end{eqnarray}
and with the material contrast parameters $\Delta\alpha^{-1}(t)$ and $\Delta\beta^{-1}(t)$ defined as
\begin{eqnarray}
\Delta\alpha^{-1}(t)&=&\alpha^{-1}(t)-\alpha_A^{-1}(t),\\
\Delta\beta^{-1}(t)&=&\beta^{-1}(t)-\beta_A^{-1}(t).
\end{eqnarray}
Equation (\ref{eq65kbl}) is the representation of the space-convolution type for a homogeneous time-dependent medium in the ${\bf k},t$-domain. It is the counterpart of the  
representation of the time-convolution type for an inhomogeneous time-independent medium \citep{Knopoff56JASA, Hoop58PHD, Gangi70JGR, Pao76JASA}.
The left-hand side represents the wave field in the actual medium at time instant $t'$.
The first term on the right-hand side describes the contributions from the fields at $t_b$ and $t_e$ (when $t'$ is between $t_b$ and $t_e$ then only the field at $t_b$ contributes).
This term is the counterpart of the spatial Kirchhoff-Helmholtz integral \citep{Morse53Book, Berkhout85Book, Frazer85GJRAS, Druzhinin98PAG, Tygel2000IP}, which is an integral along an enclosing boundary
(since time is one-dimensional, here the ``enclosing boundary'' only consists of the time instants $t_b$ and $t_e$).
If there were no contrasts of material parameters and no sources  between $t_b$ and $t_e$, then this first term would give the complete solution (similar as the Kirchhoff-Helmholtz integral would give the
complete solution if there were no contrasts and no sources inside the enclosing boundary). The second term on the right-hand side accounts for contrasts between 
the material parameters of the actual medium and those of the medium in which the Green's function is defined. 
Hence, this integral describes  wave scattering at parameter contrasts in time-dependent materials. The third term on the right-hand side accounts for the contribution of sources between
$t_b$ and $t_e$.

We transform equation (\ref{eq65kbl}) to the ${\bf x},t$-domain, using the inverse Fourier transformation defined in equation (\ref{eq52}).
Using equations (\ref{eq457}) and (\ref{eq458}) and the shift-invariance of the Green's functions, we obtain
\begin{eqnarray}
&&\hspace{-.7cm}\chi(t') U({\bf x}',t')=
\int_{\mathbb{R}^2}\Bigl( {\cal G}({\bf x}',{\bf x},t',t) {\bf\nabla}\cdot{\bf V}({\bf x},t)+\beta_A(t)\{\partial_t {\cal G}({\bf x}',{\bf x},t',t)\} U({\bf x},t)\Bigr){\rm d}{\bf x}\Bigr|_{t=t_b}^{t=t_e}\nonumber\\
&&\hspace{-.7cm}+\int_{t_b}^{t_e}\int_{\mathbb{R}^2}\Bigl( {\cal G}({\bf x}',{\bf x},t',t)\Delta\alpha^{-1}(t) {\bf \nabla}^2U({\bf x},t)
+\beta_A(t)\{\partial_t {\cal G}({\bf x}',{\bf x},t',t)\}\Delta\beta^{-1}(t) {\bf\nabla}\cdot{\bf V}({\bf x},t)\Bigr){\rm d}{\bf x}{\rm d}t\nonumber\\
&&\hspace{-.7cm}-\int_{t_b}^{t_e}\int_{\mathbb{R}^2}\Bigl(\{\partial_t {\cal G}({\bf x}',{\bf x},t',t)\}\beta_A(t) a({\bf x},t)+{\cal G}({\bf x}',{\bf x},t',t) {\bf\nabla}\cdot{\bf b}({\bf x},t)\Bigr){\rm d}{\bf x}{\rm d}t.
\label{eq65kblsp}
\end{eqnarray}
Equation (\ref{eq65kblsp}) is the representation of the space-convolution type for a homogeneous time-dependent medium in the ${\bf x},t$-domain. 
Note that equation (\ref{eq31kbl}), with the source function $s$ defined in equation (\ref{eqsource}), 
is obtained as a special case of this representation if we choose the same material parameters in both states, replace $t_b$ and $t_e$ by $-\infty$ and $\infty$, respectively,
and apply integration by parts to replace $-\int (\partial_t{\cal G})\beta a{\rm d}t$ by $\int {\cal G}\partial_t(\beta a){\rm d}t$. In its general form, equation (\ref{eq65kblsp}) provides a basis for
the analysis of wave scattering problems in homogeneous time-dependent media. We discuss an example in the next subsection. 

\subsection{Wave field representation for the case of a single time boundary}\label{sec7.2b}

We use the representation of equation (\ref{eq65kblsp}) to analyze the wave field after a time boundary between two time-independent slabs,
as illustrated in the numerical example in section \ref{sec4a}.
To this end we take $t_b$ equal to the time boundary (i.e., $t_b=50\,\mu$s) and $t_e\to\infty$. 
Hence, the boundary $t=t_e$ in the first term on the right-hand side of equation (\ref{eq65kblsp}) gives no contribution.
The wave field $U({\bf x},t)$ is the response to a spatial source distribution $s_0({\bf x})$ at $t=0$ in a 
medium with velocity $c_1=1500$ m/s for $t<t_b$ and velocity $c_2=2500$ m/s for $t>t_b$.
 The parameter $\beta$ is  constant in the numerical example, but here we assume it has the values $\beta_1$ and $\beta_2$ before and after the time boundary, respectively
 (and consequently $\alpha_1=1/\beta_1c_1^2$ and $\alpha_2=1/\beta_2c_2^2$ before and after the time boundary).
We define the Green's function in equation (\ref{eq65kblsp}) in a time-independent medium with velocity $c_A=c_2$ and parameter $\beta_A=\beta_2$ (and $\alpha_A=1/\beta_2c_2^2$)
and call this ${\cal G}_2({\bf x}',{\bf x},t',t)$. 
Note that for all $t>t_b$ we thus have $\Delta\alpha^{-1}(t)=\Delta\beta^{-1}(t)=0$, hence, the second term on the right-hand side of equation (\ref{eq65kblsp}) vanishes.
Since the source distribution  for $U({\bf x},t)$ exists only at $t=0<t_b$,  the third term on the right-hand side of equation (\ref{eq65kblsp}) also vanishes. Hence, for  $t'>t_b$ we are left with
\begin{eqnarray}
&&\hspace{-.7cm}U({\bf x}',t')=
-\int_{\mathbb{R}^2}\Bigl[ {\cal G}_2({\bf x}',{\bf x},t',t) {\bf\nabla}\cdot{\bf V}({\bf x},t)+\beta_2\{\partial_t {\cal G}_2({\bf x}',{\bf x},t',t)\} U({\bf x},t)\Bigr]_{t=t_b}{\rm d}{\bf x}.\label{eq9076}
\end{eqnarray}
In the ${\bf k},t$-domain this becomes
\begin{eqnarray}
&&\hspace{-.7cm}\check U({\bf k},t')=
-\Bigl[\check {\cal G}_2({\bf k},{\bf 0},t',t)\check V_{\rm d}({\bf k},t)+\beta_2\{\partial_t\check {\cal G}_2({\bf k},{\bf 0},t',t)\}\check U({\bf k},t)\Bigr]_{t=t_b},\quad\mbox{for}\quad t'>t_b.\label{eq9077}
\end{eqnarray}
For $t=t_b$ and $t'>t_b$ we have, according to equation (\ref{eq93bebl}),
\begin{eqnarray}
\check{\cal G}_2({\bf k},{\bf 0},t',t_b)&=&\frac{\sin(|{\bf k}|c_2(t'-t_b))}{\eta_2|{\bf k}|},\label{eq988}\\
\beta_2\partial_t\check{\cal G}_2({\bf k},{\bf 0},t',t)|_{t=t_b}&=&-\cos(|{\bf k}|c_2(t'-t_b)),
\end{eqnarray}
with $\eta_2=\beta_2c_2$. From equations (\ref{eq31kblconv}) and (\ref{eq51}), we further obtain for $t=t_b$
\begin{eqnarray}
\check U({\bf k},t_b)&=&\frac{\sin(|{\bf k}|c_1t_b)}{\eta_1|{\bf k}|}\check s_0({\bf k}),
\end{eqnarray}
with $\eta_1=\beta_1c_1$ and, using equations  (\ref{eq457}) and  (\ref{eq53}) with $\check a({\bf k},t_b)=0$,
\begin{eqnarray}
\check V_{\rm d}({\bf k},t_b)&=&-\cos(|{\bf k}|c_1t_b)\check s_0({\bf k}).
\end{eqnarray}
Substituting these expressions into equation (\ref{eq9077}) yields
\begin{eqnarray}
&&\hspace{-.7cm}\check U({\bf k},t')=\biggl(\frac{\sin \gamma_1\cos \gamma_2}{\eta_1|{\bf k}|} + \frac{\cos \gamma_1\sin \gamma_2}{\eta_2|{\bf k}|}\biggr)\check s_0({\bf k}),\quad\mbox{for}\quad t'>t_b,\label{eq90712}
\end{eqnarray}
with 
\begin{eqnarray}
\gamma_1&=&|{\bf k}|c_1t_b,\\
\gamma_2&=&|{\bf k}|c_2(t'-t_b).
\end{eqnarray}
Equation (\ref{eq90712}) can be rewritten as
\begin{eqnarray}
&&\hspace{-.7cm}\check U({\bf k},t')=\biggl(T_u\frac{\sin(\gamma_1+\gamma_2)}{\eta_1|{\bf k}|} + R_u\frac{\sin(\gamma_1-\gamma_2)}{\eta_1|{\bf k}|}\biggr)\check s_0({\bf k}),\quad\mbox{for}\quad t'>t_b,\label{eq90715}
\end{eqnarray}
with $T_u$ and $R_u$ defined in equations (\ref{eq90312}) and (\ref{eq90311}), respectively 
(this is  verified by substituting $\sin(\gamma_1\pm\gamma_2)=\sin \gamma_1 \cos \gamma_2 \pm \cos \gamma_1 \sin \gamma_2$  into equation (\ref{eq90715}), and using equations (\ref{eq48a}) and (\ref{eq48b})).
Let us write
\begin{eqnarray}
\gamma_1+\gamma_2&=&|{\bf k}|c_2(t'-t_V),\quad\mbox{with}\quad t_V=\bigl(1-\frac{c_1}{c_2}\bigr)t_b,\label{eq90716}\\
\gamma_1-\gamma_2&=&|{\bf k}|c_2(t_F-t'),\quad\mbox{with}\quad t_F=\bigl(1+\frac{c_1}{c_2}\bigr)t_b.\label{eq90717}
\end{eqnarray}
First, consider equation (\ref{eq90716}). Since $t'>t_b$ and $t_V<t_b$, we have $t'>t_V$ and, consequently, $\gamma_1+\gamma_2>0$. Hence, for the first term between the large brackets in equation (\ref{eq90715}) we find,
using equation (\ref{eq93bebl}),
\begin{eqnarray}
T_u\frac{\sin(\gamma_1+\gamma_2)}{\eta_1|{\bf k}|}=\frac{\eta_2}{\eta_1}T_u\frac{\sin(|{\bf k}|c_2(t'-t_V))}{\eta_2|{\bf k}|}
=\frac{\eta_2}{\eta_1}T_u\check{\cal G}_2({\bf k},{\bf 0},t',t_V),\quad\mbox{for}\quad t'>t_b.\label{eq90718}
\end{eqnarray}
Next, consider equation (\ref{eq90717}). Since $t'>t_b$ and $t_F>t_b$, the term $\gamma_1-\gamma_2$ can take positive as well as negative values. 
For the second term between the large brackets in equation (\ref{eq90715}) we find, using equations (\ref{eq93bebl}) and (\ref{eq1022taqbl}),
\begin{eqnarray}
R_u\frac{\sin(\gamma_1-\gamma_2)}{\eta_1|{\bf k}|}&=&\frac{\eta_2}{\eta_1}R_u\Bigl(H(t_F-t')+H(t'-t_F)\Bigr)\frac{\sin(|{\bf k}|c_2(t_F-t'))}{\eta_2|{\bf k}|}\nonumber\\
&=&\frac{\eta_2}{\eta_1}R_u\Bigl(\check{\cal G}_2^a({\bf k},{\bf 0},t',t_F)-\check{\cal G}_2({\bf k},{\bf 0},t',t_F)\Bigr),\quad\mbox{for}\quad t'>t_b.\label{eq90719}
\end{eqnarray}
Substituting equations (\ref{eq90718}) and (\ref{eq90719}) into equation (\ref{eq90715}) and  applying an inverse spatial Fourier transform yields
\begin{eqnarray}
U({\bf x}',t')=\frac{\eta_2}{\eta_1}\int_{\mathbb{R}^2}\biggl[T_u{\cal G}_2({\bf x}',{\bf x},t',t_V)+R_u\Bigl({\cal G}_2^a({\bf x}',{\bf x},t',t_F)-{\cal G}_2({\bf x}',{\bf x},t',t_F)\Bigr)\biggr]s_0({\bf x}){\rm d}{\bf x},\label{eq90720}
\end{eqnarray}
for $t'>t_b$.  The first Green's function on the right-hand side describes the transmitted, outward propagating wave in Figures \ref{Fig1} and \ref{Fig2} for $t'>t_b=50\,\mu$s in a material with 
propagation velocity $c_2$ and parameter $\beta_2$. It originates from a
virtual source at time instant $t_V$, defined in equation (\ref{eq90716}), hence, at $t_V=20\,\mu$s. The  amplitude of this transmitted wave is proportional to the transmission coefficient $T_u=0.8$
and a factor $\eta_2/\eta_1$ to account for the fact that the actual source is situated in a material with propagation velocity $c_1$ and parameter $\beta_1$.
The second Green's function is acausal and describes the reflected, inward propagating wave in Figures \ref{Fig1} and \ref{Fig2} for  $t_b<t'<t_F$. It focuses at time
instant $t_F$, defined in equation (\ref{eq90717}), hence, at $t_F=80\,\mu$s. The  amplitude of this reflected wave is proportional to the reflection coefficient $R_u=0.2$ and the factor $\eta_2/\eta_1$.
The third Green's function is  causal and describes the continuation of the reflected wave in Figures \ref{Fig1} and \ref{Fig2} for $t'>t_F=80\,\mu$s, i.e., after focusing. It originates from a
virtual source at the focal time $t_F=80\,\mu$s. The reflected wave undergoes a sign change at the focal time, hence, its amplitude after focusing is proportional to $-R_u=-0.2$ 
and the factor $\eta_2/\eta_1$.
The integration over ${\bf x}$ in equation (\ref{eq90720}) describes the spatial convolution of the superposition of these three Green's functions  with the spatial source distribution $s_0({\bf x})$ at $t=0$.
 This explains Figure \ref{Fig3} for $t'>t_b=50\,\mu$s, and in particular the recovery of the source distribution $s_0({\bf x})$ around the focal time $t_F=80\,\mu$s, where ${\cal G}_2^a({\bf x}',{\bf x},t',t_F)$ focuses.

\section{Green's function retrieval}\label{sec8}

For inhomogeneous time-independent media, it has been shown that, under specific circumstances, 
the time-correlation of passive wave measurements at two receivers converges to the response at one of these receivers as if there were
an impulsive source at the position of the other 
\citep{Weaver2001PRL, Campillo2003Science, Wapenaar2003GEO, Snieder2004PRE, Malcolm2004PRE, Roux2004JASA2, Haney2009GRL, Sabra2007APL}.
In other words, the Green's function between two receivers is retrieved by time-correlating passive responses at these receivers.
Using a reciprocity theorem of the  time-correlation type \citep{Wapenaar2006GEO}, it has been shown that this holds for arbitrary inhomogeneous time-independent media.
Here we use the reciprocity theorem of the space-correlation type of equation (\ref{eq1017tcbl}) to derive an expression for Green's function retrieval in 
a homogeneous time-dependent material  with piecewise continuous parameters $\alpha(t)$ and $\beta(t)$. 
To this end, we start with acausal Green's functions in both states, with unit sinks at $t_A$ and $t_B$, both between $t_b$ and $t_e$
(here $t_b$ and $t_e$ are again arbitrary and not related to a time boundary, such as $t_b$ in section \ref{sec7.2b}).
Substituting the quantities of Table \ref{table4} into the reciprocity theorem of the space-correlation type (equation (\ref{eq1017tcbl})), 
using reciprocity relation (\ref{eq1022taqbl}) for the Green's functions with ${\bf x}_A={\bf x}_B={\bf 0}$, yields

\begin{table}
\caption{
Quantities for deriving the expression of equation (\ref{eq1052altbl}) for Green's function retrieval.}\label{table4}
\begin{tabular}
{lll}
\hline
& {\rm State $A$} & {\rm State $B$} \\
\hline
Material parameters & $\alpha_A=\alpha(t)$ & $\alpha_B=\alpha(t)$  \\
 & $\beta_A=\beta(t)$ & $\beta_B=\beta(t)$  \\
Wave fields & $\check U_A=\check {\cal G}^a({\bf k},{\bf 0},t,t_A)$ & $\check U_B=\check {\cal G}^a({\bf k},{\bf 0},t,t_B)$\\
& $\check V_{{\rm d},A}=-\beta(t)\partial_t\check {\cal G}^a({\bf k},{\bf 0},t,t_A)$ & $\check V_{{\rm d},B}=-\beta(t)\partial_t\check {\cal G}^a({\bf k},{\bf 0},t,t_B)$\\
Sources & $\check a_A=0$ & $\check a_B=0$ \\
 & $\check b_{{\rm d},A}=-\delta(t-t_A)$ & $\check b_{{\rm d},B}=-\delta(t-t_B)$ \\
\hline
\end{tabular}
\end{table}

\begin{eqnarray}
&&\hspace{-.6cm}\check {\cal G}({\bf k},{\bf 0},t_B,t_A)-\{\check {\cal G}^a({\bf k},{\bf 0},t_B,t_A)\}^*=\nonumber\\
&&\hspace{1cm}\beta(t_b)\Bigl[\check {\cal G}^*({\bf k},{\bf 0},t_A,t)\partial_t\check {\cal G}({\bf k},{\bf 0},t_B,t) - \{\partial_t\check {\cal G}^*({\bf k},{\bf 0},t_A,t)\}\check {\cal G}({\bf k},{\bf 0},t_B,t)\Bigr]_{t=t_b}.\label{eq1052altbl}
\end{eqnarray}
Applying the inverse Fourier transform of equation (\ref{eq52}) to both sides of this equation yields
\begin{eqnarray}
&&\hspace{-.6cm}{\cal G}({\bf x}',{\bf 0},t_B,t_A)-{\cal G}^a(-{\bf x}',{\bf 0},t_B,t_A)=\nonumber\\
&&\hspace{-.2cm}\beta(t_b)\int_{\mathbb{R}^2}\Bigl[ {\cal G}({\bf x},{\bf 0},t_A,t)\partial_t {\cal G}({\bf x}'+{\bf x},{\bf 0},t_B,t) 
- \{\partial_t {\cal G}({\bf x},{\bf 0},t_A,t)\} {\cal G}({\bf x}'+{\bf x},{\bf 0},t_B,t)\Bigr]_{t=t_b}{\rm d}{\bf x}.\label{eq1052altxbl}
\end{eqnarray}
The right-hand side consists of space correlations of responses to an impulsive source at $t_b$ and its derivative, observed by receivers at $t_A$ and $t_B$. 
The left-hand side consists of a Green's function and its acausal version for a source or sink at $t_A$, observed at $t_B$. When $t_B$ is larger than $t_A$ the causal Green's function
${\cal G}({\bf x}',{\bf 0},t_B,t_A)$ is retrieved (since the acausal Green's function vanishes for this situation).
On the other hand, the acausal Green's function $-{\cal G}^a(-{\bf x}',{\bf 0},t_B,t_A)$ is retrieved when $t_B$ is smaller than $t_A$.
Note that equation (\ref{eq1052altxbl}) holds for any time instant $t_b$ of the original source, as long $t_A$ and $t_B$ are both larger than $t_b$.
Contrary to Green's function retrieval by time correlation in space-dependent media, which requires sources on a boundary enclosing the receivers, 
for Green's function retrieval by space correlation in time-dependent media it suffices to have a single source at $t_b$, prior to the receivers at $t_A$ and $t_B$.

We confirm equation (\ref{eq1052altbl}) for the situation of a time-independent medium.
Assuming $t_A>t_b$ and $t_B>t_b$, the Green's functions on the right-hand side for a time-independent medium read, according to equation (\ref{eq93bebl}),
\begin{eqnarray}
\check {\cal G}({\bf k},{\bf 0},t_A,t_b) &=& \frac{\sin(|{\bf k}|c\Delta t_A)}{\eta|{\bf k}|},\\
\partial_t\check {\cal G}({\bf k},{\bf 0},t_B,t)|_{t=t_b} &=&- \frac{1}{\beta}\cos(|{\bf k}|c\Delta t_B)
\end{eqnarray}
and similar expressions for the other Green's functions, with
\begin{eqnarray}
\Delta t_A&=&t_A-t_b,\\
\Delta t_B&=&t_B-t_b.
\end{eqnarray}
Hence, for the right-hand side of equation (\ref{eq1052altbl}) we obtain
\begin{eqnarray}
 -\frac{\sin(|{\bf k}|c\Delta t_A)\cos(|{\bf k}|c\Delta t_B)-\cos(|{\bf k}|c\Delta t_A)\sin(|{\bf k}|c\Delta t_B)}{\eta|{\bf k}|}
  = \frac{\sin\bigl(|{\bf k}|c(t_B-t_A)\bigr)}{\eta|{\bf k}|}.\label{eqA8altbl}
\end{eqnarray}
For $t_B>t_A$ this is equal to $\check {\cal G}({\bf k},{\bf 0},t_B,t_A)$,
which is the left-hand side of equation (\ref{eq1052altbl}), since $\check {\cal G}^a({\bf k},{\bf 0},t_B,t_A)=0$ for this situation.
On the other hand, for $t_B<t_A$ it is equal to $-\check {\cal G}^a({\bf k},{\bf 0},t_B,t_A)$.
This is again the left-hand side of equation (\ref{eq1052altbl}), since $\check {\cal G}^a({\bf k},{\bf 0},t_B,t_A)$ is real-valued and $\check {\cal G}({\bf k},{\bf 0},t_B,t_A)=0$ for this situation.

\section{Matrix-vector wave equation}\label{sec9}

Many authors use a matrix-vector formalism to conveniently analyse wave propagation and scattering in time-dependent media
\citep{Salem2015arXiv, Torrent2018PRB, Pacheco2020NP, Caloz2020IEEE1, Wapenaar2024WM}.
Here we recast equations (\ref{eq49}) and (\ref{eq50}) into a matrix-vector wave equation in the ${\bf x},t$-domain.
Unlike for the 1D situation, where equations similar to (\ref{eq49}) and (\ref{eq50}) 
govern the scalar wave fields $U$ and $V$ \citep{Wapenaar2024WM}, here these equations govern  $U$ and ${\bf V}$, where ${\bf V}$ is a vectorial
quantity.  Analogous to equation (\ref{eq457}), we define a scalar wave field $V_{\rm d}$ as
\begin{eqnarray}
V_{\rm d}&=&{\bf \nabla}\cdot{\bf V},\label{eq457b}
\end{eqnarray}
where subscript ${\rm d}$ refers to ``divergence''.  Using this in equation (\ref{eq49}) and applying the divergence operator to both sides of equation (\ref{eq50}) yields
\begin{eqnarray}
\partial_t U + \frac{1}{\beta}V_{\rm d}&=& a,\label{eq49b}\\
\partial_t V_{\rm d}+ \frac{1}{\alpha}{\bf \nabla}^2 U &=& b_{\rm d},\label{eq50b}
\end{eqnarray}
where, analogous to equation (\ref{eq458}), 
\begin{eqnarray}
b_{\rm d}&=&{\bf \nabla}\cdot{\bf b}.\label{eq458b}
\end{eqnarray}
Equations (\ref{eq49b}) and (\ref{eq50b}) for the scalar fields $U$ and $V_{\rm d}$ can be combined into the following matrix-vector wave equation 
\begin{eqnarray}
\partial_t{\bf q}_{\rm d}={\bf A}_{\rm d}{\bf q}_{\rm d}+{\bf d}_{\rm d},\label{eq67blue}
\end{eqnarray}
with wave field vector ${\bf q}_{\rm d}({\bf x},t)$, matrix ${\bf A}_{\rm d}({\bf x},t)$ and source vector ${\bf d}_{\rm d}({\bf x},t)$ defined as
\begin{eqnarray}
{\bf q}_{\rm d}=\begin{pmatrix}   U\\   V_{\rm d}\end{pmatrix},\quad
{\bf A}_{\rm d}=\begin{pmatrix} 0 & -\frac{1}{\beta}\\-\frac{1}{\alpha}{\bf \nabla}^2& 0 \end{pmatrix},\quad
{\bf d}_{\rm d}=\begin{pmatrix}   a\\   b_{\rm d}\end{pmatrix}.\label{eq68blue}
\end{eqnarray}
The subscript ${\rm d}$ in ${\bf q}_{\rm d}$, ${\bf A}_{\rm d}$ and ${\bf d}_{\rm d}$ is used to distinguish these quantities from those in the 1D situation where, for example, wave field vector ${\bf q}$
contains the scalar wave fields $U$ and $V$ \citep{Wapenaar2024WM}.
Equation (\ref{eq67blue}) holds for a homogeneous time-dependent medium with continuous parameters $\alpha(t)$ and $\beta(t)$.
When the material contains time boundaries, the boundary condition states that  ${\bf q}_{\rm d}({\bf x},t)$ is continuous over those time boundaries.
Equation (\ref{eq67blue}) is the counterpart of a matrix-vector wave equation for an inhomogeneous time-independent material 
\citep{Corones75JMAA, Kosloff83GEO, Fishman84JMP, Ursin83GEO, Wapenaar86GP2, Loseth2007GJI}.

Applying the 2D Fourier transformation defined in equation (\ref{eq51}) to equation (\ref{eq67blue}), we obtain the following matrix-vector wave equation in the ${\bf k},t$-domain
\begin{eqnarray}
\partial_t\check{\bf q}_{\rm d}=\check{\bf A}_{\rm d}\check{\bf q}_{\rm d}+\check{\bf d}_{\rm d},\label{eq67bl}
\end{eqnarray}
with wave field vector $\check{\bf q}_{\rm d}({\bf k},t)$, matrix $\check{\bf A}_{\rm d}({\bf k},t)$ and source vector $\check{\bf d}_{\rm d}({\bf k},t)$ defined as
\begin{eqnarray}
\check{\bf q}_{\rm d}=\begin{pmatrix}  \check U\\  \check V_{\rm d}\end{pmatrix},\quad
\check{\bf A}_{\rm d}=\begin{pmatrix} 0 & -\frac{1}{\beta}\\\frac{1}{\alpha}|{\bf k}|^2& 0 \end{pmatrix},\quad
\check{\bf d}_{\rm d}=\begin{pmatrix}  \check a\\  \check b_{\rm d}\end{pmatrix},\label{eq68bl}
\end{eqnarray}
with $\check V_{\rm d}$ and $\check b_{\rm d}$ defined in equations (\ref{eq457}) and (\ref{eq458}), respectively. Note that $\check{\bf A}_{\rm d}$ obeys the symmetry property 
\begin{eqnarray}
\check{\bf A}_{\rm d}^t{\bf N}=-{\bf N}\check{\bf A}_{\rm d},\label{eq213}
\end{eqnarray}
where superscript $t$ denotes transposition and where
\begin{eqnarray}
{{\bf N}}=\begin{pmatrix} 0 & 1 \\ -1 & 0 \end{pmatrix}.\label{eq910}
\end{eqnarray}

\section{Propagator matrix}\label{sec10}

\subsection{Propagator matrix in space-time domain}\label{sec101}

We define the propagator matrix ${\bf W}_{\rm d}({\bf x},t,t_0)$ as the solution of equation (\ref{eq67blue}) without the source term, hence
\begin{eqnarray}
\partial_t{\bf W}_{\rm d}={\bf A}_{\rm d}{\bf W}_{\rm d},\label{eq217blue}
\end{eqnarray}
with initial condition
\begin{eqnarray}
{\bf W}_{\rm d}({\bf x},t_0,t_0)={\bf I}\delta({\bf x}),\label{eq218blue}
\end{eqnarray}
where ${\bf I}$ is the identity matrix \citep{Torrent2018PRB, Salem2015arXiv, Pacheco2020NP}.
We obtain a simple representation for ${\bf q}_{\rm d}({\bf x},t)$ in terms of ${\bf W}_{\rm d}({\bf x},t,t_0)$, 
assuming they both reside in the same medium and assuming there are no sources for ${\bf q}_{\rm d}({\bf x},t)$ between $t_0$ and $t$.
Whereas ${\bf q}_{\rm d}({\bf x},t)$ can have any space-dependency at $t=t_0$, ${\bf W}_{\rm d}({\bf x},t,t_0)$ collapses to ${\bf I}\delta({\bf x})$ at $t=t_0$.
Hence, we obtain the following representation for ${\bf q}_{\rm d}({\bf x},t)$ by applying Huygens' superposition principle, according to
\begin{eqnarray}
{\bf q}_{\rm d}({\bf x},t)=\int_{\mathbb{R}^2}{\bf W}_{\rm d}({\bf x}-{\bf x}_0,t,t_0){\bf q}_{\rm d}({\bf x}_0,t_0){\rm d}{\bf x}_0,\label{eq219blue}
\end{eqnarray}
where ${\bf x}_0$ is a variable.
According to this equation, ${\bf W}_{\rm d}({\bf x},t,t_0)$ propagates the wave field vector ${\bf q}_{\rm d}$ from $t_0$ to $t$ (where $t$ can be larger or smaller than $t_0$). 
The subscript ${\rm d}$ in ${\bf W}_{\rm d}$ denotes that this propagator matrix acts on a wave field vector containing $U$ and $V_{\rm d}$, 
which is different from the 1D version of the propagator matrix, which acts on a wave field vector containing $U$ and $V$ \citep{Wapenaar2024WM}.
The propagator matrix ${\bf W}_{\rm d}({\bf x},t,t_0)$ for a homogeneous time-dependent material is the counterpart of a propagator matrix for an inhomogeneous time-independent 
material \citep{Thomson50JAP, Haskell53BSSA, Gilbert66GEO, Kennett83Book, Kennett72GJRAS, Woodhouse74GJR}.

We partition ${\bf W}_{\rm d}({\bf x},t,t_0)$ as follows
\begin{eqnarray}
{\bf W}_{\rm d}({\bf x},t,t_0)=
\begin{pmatrix} W_{\rm d}^{U,U} &  W_{\rm d}^{U,V} \\ & \\  W_{\rm d}^{V,U} &  W_{\rm d}^{V,V} \end{pmatrix}({\bf x},t,t_0).\label{eq104b}
\end{eqnarray}
The first superscript refers to the wave field quantities in ${\bf q}_{\rm d}({\bf x},t)$ in equation (\ref{eq219blue}) and the second superscript to those in ${\bf q}_{\rm d}({\bf x}_0,t_0)$ 
(with superscript $V$ referring to wave field quantity $V_{\rm d}$).

By applying equation (\ref{eq219blue}) recursively, we find the following recursive expression for ${\bf W}_{\rm d}$
\begin{eqnarray}
{\bf W}_{\rm d}({\bf x},t_N,t_0)={\bf W}_{\rm d}({\bf x},t_N,t_{N-1})*_x\cdots*_x{\bf W}_{\rm d}({\bf x},t_n,t_{n-1})*_x\cdots*_x{\bf W}_{\rm d}({\bf x},t_1,t_0),\label{eq105k}
\end{eqnarray}
where $*_x$ denotes a 2D spatial convolution, similar as in equation (\ref{eq219blue}), and where
 $t_1\cdots t_n \cdots t_{N-1}$ are time instants where the medium parameters $\alpha(t)$ and $\beta(t)$ may be discontinuous.
 Between these time instants, the parameters $\alpha(t)$ and $\beta(t)$ can, in general, be continuous functions of $t$. 
 In Appendix \ref{sec11e} we give an explicit expression for ${\bf W}_{\rm d}({\bf x},t_n,t_{n-1})$, assuming a time-independent slab between $t_{n-1}$ and $t_n$.

\begin{figure}
\centerline{\hspace{4.8cm}\epsfysize=7.5cm \epsfbox{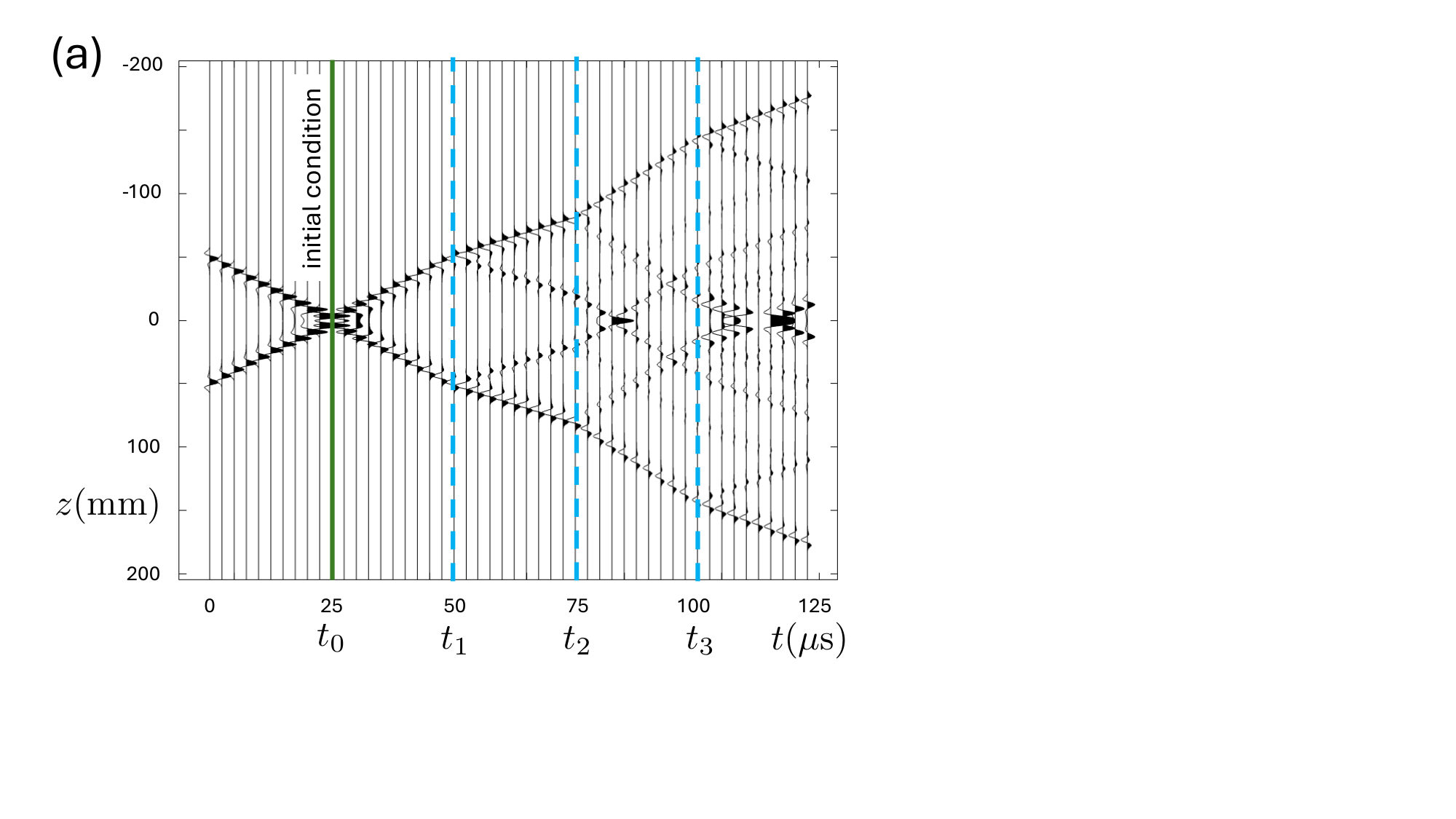}\,\hspace{-5.7cm}\epsfysize=7.5 cm \epsfbox{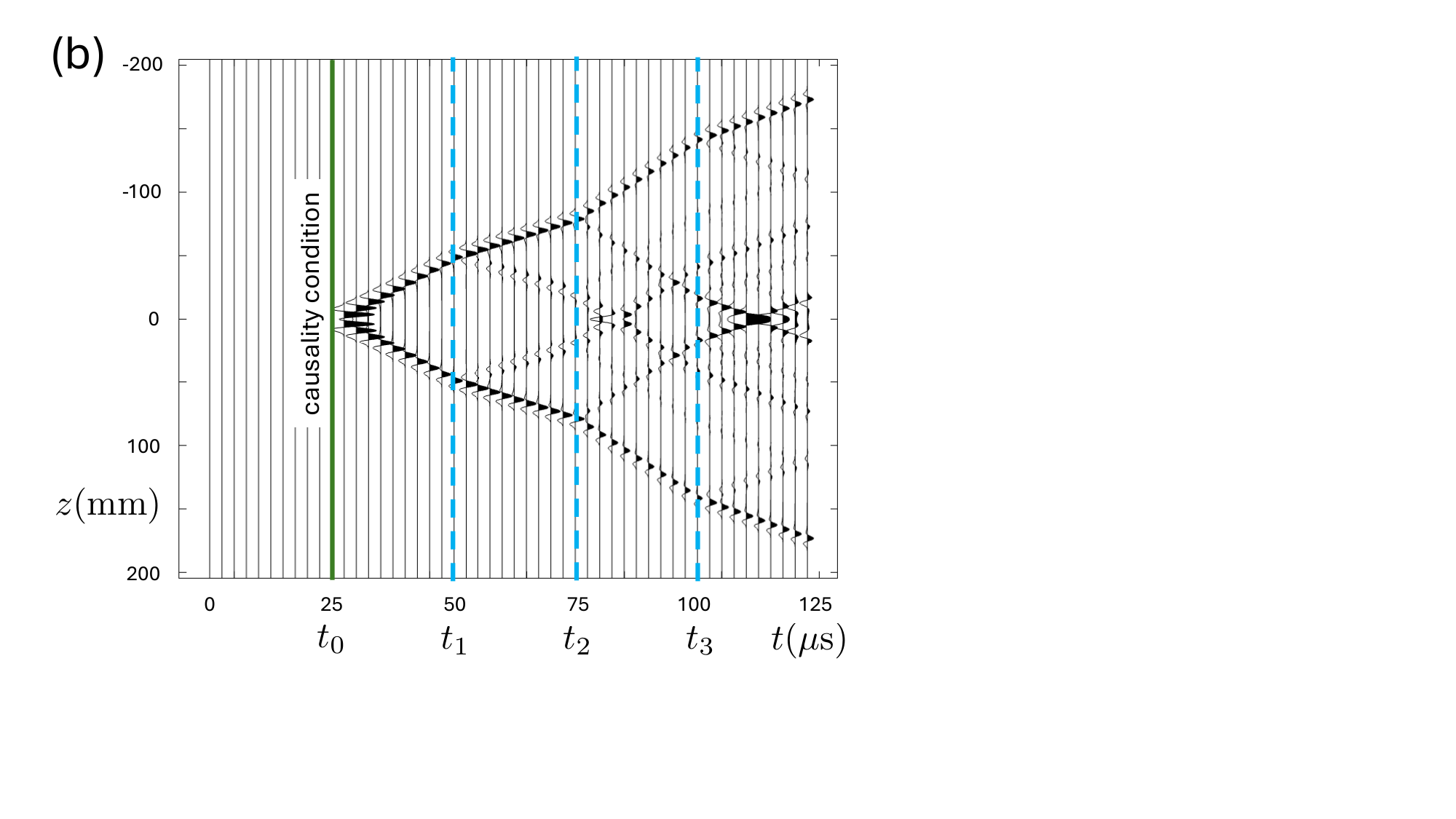}}
\vspace{-1.2cm}
\caption{Cross-sections at $x=0$ of (a) the propagator element $W_{\rm d}^{U,V}({\bf x},t,t_0)$ and (b) the Green's function  ${\cal G}({\bf x},{\bf 0},t,t_0)$
(both convolved with a spatial wavelet) for a piecewise constant time-dependent material
(for Figure (a), see also movie Wuv.mp4 in the online material).}\label{Fig4}
\end{figure}

As an illustration, we consider a piecewise constant material consisting of five time-independent slabs, with propagation velocities of
2000, 2000, 1200, 2500 and 1400 m/s, respectively. The parameter $\beta$ is taken constant. The duration of each time slab is 25 $\mu$s. 
We choose $t_0$ between the first and second time slab, i.e., $t_0=25\,\mu$s.
Figure \ref{Fig4}(a) shows a $z,t$-diagram of the propagator element $W_{\rm d}^{U,V}({\bf x},t,t_0)$ (convolved with a spatial wavelet
with a central wavenumber $k_0/2\pi=100$ m$^{-1}$) for $x=0$ and  $t_0=25\,\mu$s. 
The green line at $t_0=25\,\mu$s
indicates the initial condition of equation (\ref{eq218blue}), which for the considered off-diagonal element implies $W_{\rm d}^{U,V}({\bf x},t_0,t_0)=0$. 
In the first time slab right of this green line, the propagator  consists (for $x=0$) of upward and downward propagating direct waves. 
At each time boundary (indicated by the dashed blue lines) these waves split into upward and downward propagating transmitted and reflected waves.
This is a manifestation of the recursive character of the propagator, as formulated by equation  (\ref{eq105k}).

\subsection{Propagator matrix in wave-vector time domain}\label{sec10b}

Applying the 2D Fourier transformation defined in equation (\ref{eq51}) to equations (\ref{eq217blue}) -- (\ref{eq219blue}) we obtain
\begin{eqnarray}
\partial_t\check{\bf W}_{\rm d}=\check{\bf A}_{\rm d}\check{\bf W}_{\rm d},\label{eq217}
\end{eqnarray}
with initial condition
\begin{eqnarray}
\check{\bf W}_{\rm d}({\bf k},t_0,t_0)={\bf I}\label{eq218}
\end{eqnarray}
and representation
\begin{eqnarray}
\check{\bf q}_{\rm d}({\bf k},t)=\check{\bf W}_{\rm d}({\bf k},t,t_0)\check{\bf q}_{\rm d}({\bf k},t_0).\label{eq132}
\end{eqnarray}
Analogous to equation (\ref{eq104b}) we partition $\check{\bf W}_{\rm d}({\bf k},t,t_0)$ as
\begin{eqnarray}
\check{\bf W}_{\rm d}({\bf k},t,t_0)=
\begin{pmatrix}\check W_{\rm d}^{U,U} & \check W_{\rm d}^{U,V} \\ & \\ \check W_{\rm d}^{V,U} & \check W_{\rm d}^{V,V} \end{pmatrix}({\bf k},t,t_0).
\end{eqnarray}
By applying equation (\ref{eq132}) recursively, we find the following recursive expression for $\check{\bf W}_{\rm d}$
\begin{eqnarray}
\check{\bf W}_{\rm d}({\bf k},t_N,t_0)&=&\check{\bf W}_{\rm d}({\bf k},t_N,t_{N-1})\cdots\check{\bf W}_{\rm d}({\bf k},t_n,t_{n-1})\cdots\check{\bf W}_{\rm d}({\bf k},t_1,t_0),
\end{eqnarray}
where $t_1\cdots t_n \cdots t_{N-1}$ are time instants where the medium parameters $\alpha(t)$ and $\beta(t)$ may be discontinuous.
 Between these time instants, the parameters $\alpha(t)$ and $\beta(t)$ can, in general, be continuous functions of $t$. 
 In Appendix \ref{sec11e} we give an explicit expression for $\check{\bf W}_{\rm d}({\bf k},t_n,t_{n-1})$, assuming a time-independent slab between $t_{n-1}$ and $t_n$.

\subsection{Symmetry properties of the propagator matrix}

To derive symmetry properties of the propagator matrix, we first show that the quantity 
$\check{\bf W}_{\rm d}^t({\bf k},t,t_A){\bf N}\check{\bf W}_{\rm d}({\bf k},t,t_B)$ is a propagation invariant, i.e., that it is independent of $t$. 
Taking the time derivative, applying the product rule for differentiation, using equations (\ref{eq217}) and (\ref{eq213}), we obtain
\begin{eqnarray}
&&\partial_t\{\check{\bf W}_{\rm d}^t({\bf k},t,t_A){\bf N}\check{\bf W}_{\rm d}({\bf k},t,t_B)\}\nonumber\\
&&\hspace{2cm}=\{\partial_t\check{\bf W}_{\rm d}({\bf k},t,t_A)\}^t{\bf N}\check{\bf W}_{\rm d}({\bf k},t,t_B)+
\check{\bf W}_{\rm d}^t({\bf k},t,t_A){\bf N}\{\partial_t\check{\bf W}_{\rm d}({\bf k},t,t_B)\}\nonumber\\
&&\hspace{2cm}=\check{\bf W}_{\rm d}^t({\bf k},t,t_A)\{\check{\bf A}_{\rm d}^t{\bf N}+{\bf N}\check{\bf A}_{\rm d}\}\check{\bf W}_{\rm d}({\bf k},t,t_B)={\bf O},\label{eq1111}
\end{eqnarray}
where ${\bf O}$ is the null-matrix. This expression confirms that $\check{\bf W}_{\rm d}^t({\bf k},t,t_A){\bf N}\check{\bf W}_{\rm d}({\bf k},t,t_B)$ is independent of $t$. 
This derivation holds for continuously varying medium parameters. However, since the components of the propagator matrix are continuous over a time-boundary, 
 it follows that $\check{\bf W}_{\rm d}^t({\bf k},t,t_A){\bf N}\check{\bf W}_{\rm d}({\bf k},t,t_B)$ is a propagation
invariant for a time-dependent medium with piecewise continuous parameters. Taking $t=t_A$ and subsequently $t=t_B$, we obtain
\begin{eqnarray}
\check{\bf W}_{\rm d}^t({\bf k},t_A,t_A){\bf N}\check{\bf W}_{\rm d}({\bf k},t_A,t_B)=\check{\bf W}_{\rm d}^t({\bf k},t_B,t_A){\bf N}\check{\bf W}_{\rm d}({\bf k},t_B,t_B),\label{eq253b}
\end{eqnarray}
or, using equation (\ref{eq218}),
\begin{eqnarray}
{\bf N}\check{\bf W}_{\rm d}({\bf k},t_A,t_B)=\check{\bf W}_{\rm d}^t({\bf k},t_B,t_A){\bf N},\label{eq253}
\end{eqnarray}
or, applying the inverse 2D Fourier transformation defined in equation (\ref{eq52}),
\begin{eqnarray}
{\bf N}{\bf W}_{\rm d}({\bf x},t_A,t_B)={\bf W}_{\rm d}^t({\bf x},t_B,t_A){\bf N}.\label{eq253c}
\end{eqnarray}
Equations (\ref{eq253}) and (\ref{eq253c}) 
formulate symmetry properties of the propagator matrix in the ${\bf k},t$- and ${\bf x},t$-domain, respectively.
Substituting equations (\ref{eq910}) and (\ref{eq104b}) into equation (\ref{eq253c}) we find 
the following symmetry properties for the components of the propagator matrix 
\begin{eqnarray}
W_{\rm d}^{U,V}({\bf x},t_A,t_B)&=&-W_{\rm d}^{U,V}({\bf x},t_B,t_A),\label{eq1014}\\
W_{\rm d}^{V,U}({\bf x},t_A,t_B)&=&-W_{\rm d}^{V,U}({\bf x},t_B,t_A),\label{eq1015}\\
W_{\rm d}^{V,V}({\bf x},t_A,t_B)&=&W_{\rm d}^{U,U}({\bf x},t_B,t_A).\label{eq1016}
\end{eqnarray}
Next we show that three of the four components of the propagator matrix can be expressed in terms of the upper-right component
$W_{\rm d}^{U,V}({\bf x},t,t_0)$. From equations (\ref{eq68blue}), (\ref{eq217blue}) and (\ref{eq104b}) we find
\begin{eqnarray}
 W_{\rm d}^{V,V}({\bf x},t,t_0)&=&-\beta(t)\partial_t W_{\rm d}^{U,V}({\bf x},t,t_0),\label{eq231}\\
 W_{\rm d}^{V,U}({\bf x},t,t_0)&=&-\beta(t)\partial_t W_{\rm d}^{U,U}({\bf x},t,t_0).\label{eq230}
\end{eqnarray}
From equations (\ref{eq1016}), (\ref{eq231}) and (\ref{eq1014}) we find
\begin{eqnarray}
 W_{\rm d}^{U,U}({\bf x},t,t_0)&=&\beta(t_0)\partial_{t_0} W_{\rm d}^{U,V}({\bf x},t,t_0).\label{eq232}
\end{eqnarray}
Substitution into equation (\ref{eq230}) yields
\begin{eqnarray}
 W_{\rm d}^{V,U}({\bf x},t,t_0)&=&-\beta(t)\beta(t_0)\partial_t\partial_{t_0} W_{\rm d}^{U,V}({\bf x},t,t_0).\label{eq233}
\end{eqnarray}
Hence, equations (\ref{eq231}),  (\ref{eq232}) and  (\ref{eq233}) are the sought relations.

\subsection{Relations between the Green's function and the propagator matrix}\label{sec10d}

We show that the Green's function ${\cal G}({\bf x},{\bf 0},t,t_0)$, obeying wave equation (\ref{eq31agbl}) for ${\bf x}_0={\bf 0}$
with causality condition (\ref{eq31con}), can be expressed in terms of the upper-right component of the propagator matrix, according to
\begin{eqnarray}
 {\cal G}({\bf x},{\bf 0},t,t_0)=-H(t-t_0) W_{\rm d}^{U,V}({\bf x},t,t_0).\label{eq1021}
\end{eqnarray}
Note that this relation is different from that for the 1D situation (which includes a spatial derivative on the left-hand side \citep{Wapenaar2024WM}) 
because of the deviating definition of the wave field vector ${\bf q}_{\rm d}$ 
and consequently of the propagator matrix ${\bf W}_{\rm d}$, see sections \ref{sec9} and \ref{sec101}.

Due to the Heaviside function in equation (\ref{eq1021}), the causality condition is fulfilled. In the following we show that 
$H(t-t_0) W_{\rm d}^{U,V}({\bf x},t,t_0)$ obeys the same wave equation as $-{\cal G}({\bf x},{\bf 0},t,t_0)$.
For the first time derivative we have
\begin{eqnarray}
\partial_t\{H(t-t_0) W_{\rm d}^{U,V}({\bf x},t,t_0)\}=\delta(t-t_0) W_{\rm d}^{U,V}({\bf x},t,t_0)+H(t-t_0)\partial_t W_{\rm d}^{U,V}({\bf x},t,t_0).\label{eq1123}
\end{eqnarray}
From equations (\ref{eq218blue}) and (\ref{eq104b}) we have $W_{\rm d}^{U,V}({\bf x},t_0,t_0)=0$. Using this and equation (\ref{eq231}) in equation (\ref{eq1123}), we obtain
\begin{eqnarray}
\partial_t\{H(t-t_0) W_{\rm d}^{U,V}({\bf x},t,t_0)\}=
-\frac{1}{\beta(t)}H(t-t_0) W_{\rm d}^{V,V}({\bf x},t,t_0).
\end{eqnarray}
From equations (\ref{eq68blue}), (\ref{eq217blue}), (\ref{eq218blue}) and (\ref{eq104b}) we have
$\partial_tW_{\rm d}^{V,V}({\bf x},t,t_0)=-\frac{1}{\alpha(t)}{\bf \nabla}^2W_{\rm d}^{U,V}({\bf x},t,t_0)$ and 
$W_{\rm d}^{V,V}({\bf x},t_0,t_0)=\delta({\bf x})$. Using this, we find for the second time derivative
\begin{eqnarray}
\partial_t\bigl(\beta(t)\partial_t\{H(t-t_0) W_{\rm d}^{U,V}({\bf x},t,t_0)\}\bigr)&=&-\partial_t\{H(t-t_0) W_{\rm d}^{V,V}({\bf x},t,t_0)\}\\
&=&-\delta(t-t_0)\delta({\bf x})+\frac{1}{\alpha(t)}{\bf \nabla}^2\{H(t-t_0) W_{\rm d}^{U,V}({\bf x},t,t_0)\}.\nonumber
\end{eqnarray}
Comparing this with equation (\ref{eq31agbl}) for ${\bf x}_0={\bf 0}$, with $c(t)$ defined in equation (\ref{eq8a}),
we conclude that $H(t-t_0) W_{\rm d}^{U,V}({\bf x},t,t_0)$ indeed obeys the same wave equation as $- {\cal G}({\bf x},{\bf 0},t,t_0)$.
This completes the proof of equation (\ref{eq1021}). Figure \ref{Fig4}(b) 
shows a $z,t$-diagram of the Green's function  ${\cal G}({\bf x},{\bf 0},t,t_0)$  (convolved with a spatial wavelet) for $x=0$ and  $t_0=25\,\mu$s. 
It has been obtained from $W_{\rm d}^{U,V}({\bf x},t,t_0)$ in Figure \ref{Fig4}(a)  via equation (\ref{eq1021}).
The green line at $t_0=25\,\mu$s indicates the causality condition of equation (\ref{eq31con}).

In a similar way as above we can show that the acausal Green's function ${\cal G}^a({\bf x},{\bf 0},t,t_0)$, with condition
${\cal G}^a({\bf x},{\bf 0},t,t_0)=0$ for $t>t_0$, can be expressed as
\begin{eqnarray}
 {\cal G}^a({\bf x},{\bf 0},t,t_0)=H(t_0-t) W_{\rm d}^{U,V}({\bf x},t,t_0).\label{eq1024}
\end{eqnarray}
Vice versa, by combining equations (\ref{eq1021}) and (\ref{eq1024}), we can express $W_{\rm d}^{U,V}({\bf x},t,t_0)$ in terms of the Green's function and its acausal version, according to
\begin{eqnarray}
 W_{\rm d}^{U,V}({\bf x},t,t_0)= {\cal G}^a({\bf x},{\bf 0},t,t_0)- {\cal G}({\bf x},{\bf 0},t,t_0).\label{eq1025}
\end{eqnarray}
Using equations  (\ref{eq231}),  (\ref{eq232}) and  (\ref{eq233}), the other components of  the propagator matrix
${\bf W}_{\rm d}({\bf x},t,t_0)$ can also be expressed in terms of the Green's function and its acausal version.

Note that the simple relations between the Green's function and the propagator matrix derived here 
are specific for a homogeneous time-dependent medium. They 
cannot be translated to the situation of an inhomogeneous time-independent medium by an exchange of 
space- and time-coordinates because
the causality conditions for the Green's function are the same for both types of medium \citep{Wapenaar2024WM}.

\section{Conclusions}
We discussed fundamental aspects of wave propagation and scattering in 2D homogeneous time-dependent materials, 
using a unified notation which simultaneously captures electromagnetic, acoustic and elastodynamic shear waves.

We reviewed transmission and reflection of a plane wave that is incident on a time boundary. 
Due to causality, the transmitted and reflected waves both exist only after the time boundary. 
The net field-momentum density is conserved over a time boundary, but the net power-flux density is not.

We reviewed the Green's function of a time-dependent material, being the causal response to an impulsive point source. For the special case of
a single time boundary, the reflected part of the Green's function after the time boundary propagates back and focuses at the position of the original source.
Subsequently, the focus acts as a virtual source for a wave field with opposite amplitude.

We discussed propagation invariants (i.e., time-independent quantities) for homogeneous, piecewise continuous time-dependent materials. 
The net field-momentum density is a special case of a propagation invariant:
it is conserved not only over time boundaries (as mentioned above), but also in the continuously varying material between the time boundaries.
More general propagation invariants have been formulated for specific combinations of wave fields in two mutually independent states.
These propagation invariants form the basis for general reciprocity theorems of the space-convolution type and of the space-correlation type for homogeneous, piecewise continuous time-dependent materials.

We used the reciprocity theorem of the space-convolution type to derive a source-receiver reciprocity relation for the Green's function of a time-dependent material:
the causal Green's function between a source at $t_B$ and a receiver at $t_A$ is identical to the {\it acausal} Green's function between a sink at $t_A$ and a receiver at $t_B$.
We also used the reciprocity theorem of the space-convolution type to derive a general wave-field representation for a homogeneous time-dependent material. Similar as its counterpart 
for an inhomogeneous time-independent material, it forms the basis for the analysis of wave scattering problems. As an example, we used this representation to quantitatively explain the 
behaviour of the Green's function after a single time boundary.
We used the reciprocity theorem of the space-correlation type to derive a representation for Green's function retrieval
from passive measurements in a homogeneous, piecewise continuous time-dependent material.

Finally, we formulated a matrix-vector wave equation for time-dependent materials and used this as a basis for deriving a propagator matrix, its symmetry properties and its relation with 
the causal and acausal Green's functions.

We restricted the analysis in this paper to 2D homogeneous time-dependent materials. 
For acoustic waves (row 3 in Table \ref{table1}), almost all expressions in this paper remain valid for 3D homogeneous time-dependent materials when the vectors 
${\bf V}$, ${\bf Q}$, ${\bf b}$, ${\bf x}$, ${\bf k}$ and ${\bf \nabla}$ are extended with a $y$-component 
and all integrals over $\mathbb{R}^2$ are replaced by integrals over $\mathbb{R}^3$.
Only the right-hand side of equation (\ref{eq71bl}) needs to be replaced by $\alpha\delta\bigl(t-t_0- |{\bf x}-{\bf x}_0|/{c}\bigr)/4\pi|{\bf x}-{\bf x}_0|$ 
(and similar replacements should be made in equation (\ref{eqB3q})).
For electromagnetic and elastodynamic waves the extension to 3D is more involved and should be derived from multicomponent equations, 
such as those given by references \citep{Hoop2014WM, Salem2015arXiv}. This is left for future research. 

The analysis of wave propagation and scattering in materials that are simultaneously space- and time-dependent is not straightforward. 
Whereas effective medium theory can be used for periodic small-scale space-time variations \citep{Trainiti2016NJP, Nassar2017JMPS, Nassar2017RS, Goldsberry2019JASA, Sotoodehfar2023OE}, 
the treatment of general piecewise continuous materials with mixed
space-time boundaries and arbitrary varying space-time materials between these boundaries is more complex. One of the reasons is that 
$P$ and ${\bf Q}$ are the preferred wave field quantities to be analyzed at space boundaries and in space-dependent regions (case ii in section \ref{sec2}), 
whereas $U$ and ${\bf V}$ are the preferred quantities at time boundaries and in  time-dependent regions (case iii in section \ref{sec2}).
Many special situations are discussed in the literature. For example, 
 \cite{Caloz2020IEEE2} discusses how to model scattering at a mixed space-time boundary,
 \cite{Apffel2022PRL} shows how multiple distinct space and time boundaries can be used to control frequency conversion and 
 \cite{Manen2024arXiv} discusses the design of an acoustic space-time material (with distinct space  and time boundaries) which computes its own inverse.
A fundamental treatment of wave propagation and scattering in arbitrary space-time materials remains subject to further research.

\vskip6pt

\enlargethispage{20pt}

\section*{Data access}
Links to movies associated with Figures \ref{Fig1}, \ref{Fig3} and \ref{Fig4} and a matlab-code to reproduce the figures and movies can be found on
{\tt https://gitlab.com/geophysicsdelft/OpenSource} in the directory {\tt.../timematerial}.
\section*{Acknowledgements}
The author thanks Johannes Aichele and Dirk-Jan van Manen for the inspiring discussions about waves in time-dependent materials. 
The constructive comments of the three referees are much appreciated and helped to improve the clarity of the paper.

\appendix

\section{Inverse Fourier Transform of wave-vector time domain Green's function}\label{App2}

We evaluate the inverse Fourier transform of $\check {\cal G}({\bf k},{\bf x}_0,t,t_0)$, defined in equation (\ref{eq93bebl}). Using equation (\ref{eq52}), we write
\begin{eqnarray}
{\cal G}({\bf x},{\bf x}_0,t,t_0)&=&\frac{1}{4\pi^2} \int_{{\mathbb R}^2} \check {\cal G}({\bf k},{\bf x}_0,t,t_0)  \exp (i{\bf k}\cdot{\bf x}){\rm d}{\bf k}\nonumber\\
&=&\frac{H(\Delta t)}{4\pi^2\eta} \int_{{\mathbb R}^2} \frac{\sin(k_rc\Delta t)}{k_r} \exp (i{\bf k}\cdot\Delta{\bf x}){\rm d}{\bf k},\label{eqA1}
\end{eqnarray}
with $\Delta t=t-t_0$, $k_r=|{\bf k}|$ and $\Delta{\bf x}={\bf x}-{\bf x}_0$. Defining polar coordinates ${\bf k}=k_r(\cos\theta,\sin\theta)$ and $\Delta{\bf x}=r(\cos\phi,\sin\phi)$, with $r=|\Delta {\bf x}|$,
we have ${\bf k}\cdot\Delta{\bf x}=k_rr\cos(\theta-\phi)$ and ${\rm d}{\bf k}=k_r{\rm d}k_r{\rm d}\theta$, hence
\begin{eqnarray}
{\cal G}({\bf x},{\bf x}_0,t,t_0)&=&\frac{H(\Delta t)}{2\pi\eta} \int_0^\infty{\rm d}k_r \sin(k_rc\Delta t)\frac{1}{2\pi}\int_0^{2\pi}\exp(ik_rr\cos(\theta-\phi)){\rm d}\theta.
\end{eqnarray}
The integral over $\theta$ is independent of $\phi$, so we may choose $\phi=0$. Furthermore, this integral from $0$ to $2\pi$ is twice the real part of the
integral from $0$ to $\pi$. Hence
\begin{eqnarray}
{\cal G}({\bf x},{\bf x}_0,t,t_0)&=&\frac{H(\Delta t)}{2\pi\eta} \int_0^\infty{\rm d}k_r \sin(k_rc\Delta t)\frac{1}{\pi}\int_0^{\pi}\cos(k_rr\cos\theta){\rm d}\theta.
\end{eqnarray}
Using the definition of the Bessel function $J_0$ in equation (9.1.18) of \citep{Abramowitz70Book}, we obtain
\begin{eqnarray}
{\cal G}({\bf x},{\bf x}_0,t,t_0)&=&\frac{H(\Delta t)}{2\pi\eta} \int_0^\infty \sin(k_rc\Delta t)J_0(k_rr){\rm d}k_r.
\end{eqnarray}
Finally, using  equation (6.671-7) of \citep{Gradshteyn94Book} and the definitions of $\Delta t$ and $r$ given below equation (\ref{eqA1}) yields
\begin{eqnarray}
{\cal G}({\bf x},{\bf x}_0,t,t_0)&=&\frac{1}{2\pi\beta c^2}\frac{H\Bigl(t-t_0- \frac{|{\bf x}-{\bf x}_0|}{c}\Bigr)}{\sqrt{(t-t_0)^2- \frac{|{\bf x}-{\bf x}_0|^2}{c^2}}}.\label{eq71blapp}
\label{eq52app}
\end{eqnarray}

\section{Propagator matrix for a time-independent slab}\label{sec11e}

For the special case of a time-independent slab between $t_{n-1}$ and $t_n$, with parameters $\alpha_n$ and $\beta_n$,
the solution of equation (\ref{eq217}), with initial condition (\ref{eq218}) (with $t_0$ replaced by $t_{n-1}$ and $t$ by $t_n$), reads
\begin{eqnarray}
\check {\bf W}_{\rm d}({\bf k},t_n,t_{n-1})=
\begin{pmatrix}
\cos(|{\bf k}|c_n \Delta t_n) & -\frac{1}{\eta_n|{\bf k}|}\sin(|{\bf k}|c_n \Delta t_n)\\
\eta_n|{\bf k}|\sin(|{\bf k}|c_n \Delta t_n) & \cos(|{\bf k}|c_n \Delta t_n)
\end{pmatrix},
\end{eqnarray}
with $c_n=1/\sqrt{\alpha_n\beta_n}$, $\eta_n=\beta_nc_n$ and $\Delta t_n=t_n-t_{n-1}$.
Note that the upper-right component can, according to the Fourier transform of equation (\ref{eq1025}), be expressed as
\begin{eqnarray}
\check W_{\rm d}^{U,V}({\bf k},t_n,t_{n-1})=\check {\cal G}^a({\bf k},{\bf 0},t_n,t_{n-1})-\check {\cal G}({\bf k},{\bf 0},t_n,t_{n-1}),
\end{eqnarray}
with $\check {\cal G}({\bf k},{\bf 0},t_n,t_{n-1})$ given in equation (\ref{eq93bebl}) for ${\bf x}_0={\bf 0}$, $t_0=t_{n-1}$ and $t=t_n$.
Similarly, we obtain an explicit expression for  $W_{\rm d}^{U,V}({\bf x},t_n,t_{n-1})$ in the space-time domain by substituting equation 
(\ref{eq52app}) for ${\bf x}_0={\bf 0}$ into equation (\ref{eq1025}), using equation (\ref{eq1022sqshbl}).
This gives
\begin{eqnarray}
W_{\rm d}^{U,V}({\bf x},t_n,t_{n-1})&=&{\cal G}({\bf x},{\bf 0},t_{n-1},t_n)-{\cal G}({\bf x},{\bf 0},t_n,t_{n-1})\nonumber\\
&=&\frac{1}{2\pi\beta_n c_n^2}\frac{H(-\Delta t_n- |{\bf x}|/c_n)-H(\Delta t_n- |{\bf x}|/c)}{\sqrt{\Delta t_n^2- |{\bf x}|^2/c_n^2}}.\label{eqB3q}
\end{eqnarray}
Explicit expressions for the other components of ${\bf W}_{\rm d}({\bf x},t_n,t_{n-1})$ follow by substituting equation (\ref{eqB3q}) into equations  (\ref{eq231}),  (\ref{eq232}) and  (\ref{eq233}), 
with $t_0$ and $t$ replaced by $t_{n-1}$ and $t_n$, respectively.

\bibliographystyle{gji}


\end{spacing}
\end{document}